\documentclass[preprint]{aastex}
\usepackage{amsmath}
\usepackage{color}
\usepackage[title]{appendix}
\def\eg{\emph{e.g.},}

\def\etal{\emph{et al.}}
\def\cf{\emph{cf.~}}

\begin{document}
\title{Characterizing the Uncertainty in Cluster Magnetic Fields derived from Rotation Measures}
\author{A. R. Johnson\altaffilmark{1,2},  L. Rudnick\altaffilmark{2}, T. W. Jones\altaffilmark{2,3}, P. J. Mendygral\altaffilmark{2,4}, \and K. Dolag\altaffilmark{5}}
\altaffiltext{1}{Optum, Inc., 13625 Technology Dr., Eden Prairie, MN 55344}
\altaffiltext{2}{Minnesota Institute for Astrophysics, School of Physics and Astronomy, University of Minnesota, 116 Church St., SE, Minneapolis, MN 55455}
\altaffiltext{3}{Minnesota Supercomputing Institute, University of Minnesota, Minneapolis, MN 55455}
\altaffiltext{4}{Cray Inc., 2131 Lindau Lane, Suite 1000, Bloomington, MN 55425}
\altaffiltext{5}{Universit\"ats-Sternwarte, Fakult\"at f\"ur Physik, Ludwig-Maximilians Universit\"at M\"unchen, Scheinerstrasse 1, D-81679 M\"unchen, Germany, and
Max-Planck-Institut f\"ur Astrophysik, Karl-Schwarzschild Strasse 1, D-85748 Garching bei M\"unchen, Germany}
\date{\today}
\newcommand{\red}[1]{\textcolor{red}{#1}}
\newcommand{\green}[1]{\textcolor{green}{#1}}
\newcommand{\lrhere}[0]{\large{\bf \red{LR HERE}}\normalsize\\}
\newcommand{\ed}[1]{#1}
\keywords{galaxies: clusters: magnetic fields - Faraday Rotation}


\begin{abstract}
Magnetic fields play vital roles in intracluster media (ICMs), but estimating their strengths and distributions from observations is a major challenge. Faraday rotation measures (RMs) are widely applied to this task, so it is critical to understand inherent uncertainties in RM analysis. {\it{In this paper, we seek to characterize those  uncertainties given the types of information available today, independent of the specific technique used}}.  We conduct synthetic RM observations through the ICM of a galaxy cluster drawn from an MHD cosmological simulation in which the magnetic field is known. We analyze the synthetic RM observations using an analytical formalism based on commonly used model assumptions allowing us to relate model physical variables to outcome uncertainties. 
Despite the simplicity of some assumptions, and unknown physical parameters, we are able to extract an approximate magnitude of the central magnetic field within an apparently irreducible uncertain factor $\approx$ 3.  Principal, largely irreducible, uncertainties come from the unknown depth along the line of sight of embedded polarized sources, the lack of robust coherence lengths from area-constrained polarization sampling, and the unknown scaling between ICM electron density and magnetic field strength.  The RM-estimated central magnetic field strengths span more than an order of magnitude including the full range of synthetic experiments.

\end{abstract}


\section{Introduction}
The hot, diffuse media of galaxy clusters (ICMs) are  magnetized  \citep[e.g.,][]{MagFieldReview} and very likely turbulent \citep[\eg][]{cturb,cturb2,miniati15,vazza17a,vazza17b}. The
strength and structure of the magnetic fields play central roles in determining the turbulent and thermodynamical
properties of the ICMs. For instance, these properties may control the scale and isotropy of transport processes such as viscosity and thermal conduction, even
if the magnetic, Maxwell stresses are insignificant on cluster scales.  
Magnetic fields or their induced anisotropic transport characteristics can stabilize structures such as cold fronts \citep[e.g.,][]{zuhone11} or lead to instabilities that, for example, influence cluster thermal structure \citep[e.g.,][]{aniso}.  ICM magnetic fields also control the acceleration and transport of relativistic particles within the ICMs \citep[\eg][]{brun14}. For these reasons, much effort has gone into observational estimates of ICM magnetic field properties.  As discussed below, there are key physical properties of the field and of the accompanying thermal plasma that are not well constrained observationally, and therefore lead to irreducible uncertainties in the structure and strength of the derived magnetic fields.  It is the goal of the paper to characterize the most important uncertainties in this process.

Cluster magnetic fields reveal themselves through the diffuse synchrotron emission found in a good many clusters [such as giant radio halos, \citep[\eg][]{feretti12,brun14}] and through the Faraday rotation of linearly polarized radio emission propagating through the ICM. 
Faraday rotation, is based on circular birefingence, in which the plane of
polarization of a linearly polarized signal rotates along the line of sight by an amount that is proportional to the square of the wavelength of the radiation  and to the integral electron along the line of sight(i.e., $\Delta\phi$=$RM \times \lambda^2$). The constant of proportionality, the rotation measure (RM), is determined by the integral along the line of sight of the plasma electron density, $n_e$, times the projection of the vector magnetic field onto the line of sight, $B_{\parallel}$; namely over path length $\ell$,
\begin{align}
RM(\ell) = 812 \int_0^\ell \! n_{e}(s) B_{\parallel}(s) \, \mathrm{d} s ~~{\rm rad ~m^{-2}},
\label{eq:RM-def}
\end{align}
{where $n_e$ is expressed in units cm$^{-3}$, $B_{\parallel}$ is in units of $\mu$Gauss, and the differential line of sight path, $\mathrm{d} s$, is in units of kpc. We assume, going forward that foreground RM contributions (\eg ~galactic) have been removed and that we can ignore extraneous modifications to the ICM due, for example, to AGN outflows.}

RM analyses of ICMs have been applied extensively, using polarized synchrotron sources embedded in the cluster \citep[e.g.,][]{hydra1,perseus1,AbellRMs,Abell2199,Abell119,RandomLOS,guidetti08,laing08,guidetti10,A194}  as well as multiple unresolved polarized sources behind galaxy clusters. \citep[e.g.,][]{clarke99,Abell119,Bonafede10} 
RM measurements have typically suggested maximum magnetic field strengths at cluster centers 
 $\sim 1-15 \mu \mathrm{G}$, with strengths steadily decreasing outwards.
Several recent MHD cosmological simulation studies \citep[e.g.,][]{donn99,XuRMs} have obtained evolved cluster magnetic field distributions that are relatively insensitive
to the model details for the seed fields, and that are qualitatively consistent
with the reported observations.

So far, however, neither the simulated, nor the RM-based ``observed'' field distributions provide sufficiently robust information needed to constrain ICM physics and evolution. In this paper we aim to examine some of the factors that limit the accuracy of ICM RM studies, specifically targeting limitations coming from inherent, irreducible uncertainties in such analyses. In order to do this we have adopted a simple, analytic approach that cleanly exposes such limitations. We emphasize at the start that our intent is not to promote any particular analysis method, but to identify inherent model and physical uncertainties that seem to restrict accuracy to something close to an order of magnitude.

 Application of Eqn. \ref{eq:RM-def} to establish the magnetic field distribution along a given line of sight within an ICM is, in fact, not straightforward. First, it requires that one isolate the electron density distribution, $n_e(l)$, from the magnetic field distribution. That can often be done reasonably well in clusters by modeling thermal X-ray measurements. A more serious issue is revealed by the fact that measured ICM RM distributions are patchy and irregular, often including RMs of both signs, telling us that the magnetic fields are disordered. Thus, $B_{\parallel}$ contributions to Eqn. \ref{eq:RM-def} are stochastic; a statistical analysis is mandated. In practice one usually tries to estimate the dispersion of the magnetic field strength distribution, $\sigma_{B,\parallel}$.

On the reasonable and common assumption that the ICM magnetic field averaged over large volumes is isotropically disordered by turbulence with a magnetic field coherence scale, $L_{coh} \ll \ell$, with $\ell$ the full path length, an ensemble of statistically independent lines of sight lead to $\langle RM\rangle \rightarrow 0$ while the RM dispersion, $\sigma_{RM} = \sqrt{\langle RM^2\rangle - |\langle RM\rangle^2|} \rightarrow \sqrt{\langle RM^2\rangle}$. 
More to the point of the exercise, $\sigma_{RM} \propto \sigma_{B,\parallel} =\sqrt{\langle B_{\parallel}^2\rangle - \langle B_{\parallel}\rangle^2}\rightarrow \sqrt{\langle B_{\parallel}^2\rangle}\rightarrow \sqrt{\langle B^2\rangle/3}$.\ed{ In the idealized case of a homogeneous medium, and independent lines of sight, $\sigma_{RM} \propto \sqrt{L_{coh} \ell}$.} There are multiple metrics in the literature for magnetic field coherence scales, as outlined in Appendix. These include, for example, the magnetic field autocorrelation length, $L_B$, \citep{Bayesian2} and the so-called integral scale of the magnetic field, $L_{int}$, which is the power-spectrum-weighted mean length associated with the magnetic field variations \citep{ClusterLengths}.  For an isotropic, {homogeneous}, turbulent magnetic field, $L_{int} = 2 L_B$. If we choose a characteristic scale, $\Lambda = (3/2) L_B$, then the relation between $\sigma_{RM}$ and $\sigma_{B,\parallel}$ conveniently takes the familiar normalization in Eqn. \ref{eq:RM-def}; namely, \citep[e.g.,][]{lawler82,tribble91,feretti95,Felten96} 
\begin{align}
\sigma_{RM} = 812 \, \bar{n}_e \sigma_{B,\parallel} \,\sqrt{\ell\Lambda}= 812 \, \bar{n}_e \sigma_{B,\parallel} \, \Lambda \,\sqrt{\frac{\ell}{\Lambda}},
\label{eq:sigmaRM}
\end{align}
where the units are as in Eqn. \ref{eq:RM-def}. We will adopt this common convention below.  We also, for simplicity, associate $\Lambda$ with the projected, observable, RM coherence length, since that accurately applies in a homogeneous medium. Deviations from this in an ICM context, where there are large scale gradients will be addressed below. Note from the beginning that, unless the various lines of sight sample regions with uncorrelated magnetic fields, so include separations $> \Lambda$, but regions with similar electron densities, the calculated RM dispersion measure from Eqn. \ref{eq:sigmaRM} is not identical to $\sigma_{RM}$ among different lines of sight. \ed{This equivalence cannot be taken for granted in observational analyses based
on sparse RM information \citep[e.g.,][]{obsmodel1}.}

{There are a number of important challenges in applying Eqn. \ref{eq:sigmaRM} to actual data; these lead to limitations in the accuracy with which we can characterize the magnetic field strengths in clusters. These challenges include: 
\begin{itemize}
\itemsep -5pt
\item{the inhomogeneous and anisotropic magnetic field structure in realistic ICMs, as illuminated in MHD cluster formation simulations;}
\item{the unknown position of cluster-embedded polarized sources along the line of sight through the ICM;}
\item{\ed{the unknown scaling between magnetic field and thermal plasma density -- the
latter, in turn, must be modeled using X-ray or microwave observations;}}
\item{the incomplete sampling coming from limited availability of polarized emissions of appropriately large scales;}
\item{\ed{for sight lines penetrating the core, the bulk of the emission comes from a very small number of high emissivity regions, so the sampling of independent field regions is reduced;}}
\item{the unknown radial dependence of characteristic spatial scales in the ICM turbulence;}
\item{the fact that $\sigma_{RM}$ and $\Lambda$ are inherently statistical quantities, with distributions that can only be sampled, but not fully measured. \citet{newman} have argued that the inherent uncertainties in their estimation are comparable to their magnitudes;
\item{modification of the local magnetic field and density structure by radio lobe interactions with the ICM, as discussed briefly below.}}
\end{itemize}
}

The contribution of RM variations local to an embedded radio galaxy, as opposed to RM variations which characterize the unperturbed ICM, is an open issue.  \citet{bandedRMs} and \citet{AsymRMs}, e.g.,  reported anistropic RM patterns associated with several cluster AGNs, implying, they argued, influences by the AGNs on the RM distribution.  \citet{RSContribution} argued that the RM distributions observed in front of several embedded
cluster sources  showed biases suggesting influence from plasma entrained within the emitting radio lobes. \citet{RBRebuttle} argued, however,
that those biases were not statistically robust. On the theoretical front, \citet{espinosaFR2} conducted synthetic RM observations of MHD simulations of high powered, FR II AGN jets driven into a model
ICM with an initially isotropic, random magnetic field, and found that the jet interactions modified the RM statistics in ways that biased the inferred
magnetic field values upwards by as much as 70\%. The complex issue of physical modification of the local ICM interacting with a radio sources is beyond the scope of this paper and not addressed. 

{It is the purpose of this paper to use extensions of Eqn. \ref{eq:sigmaRM} in order to explore and quantify many of these uncertainties.} Our approach is to carry out synthetic RM measurements of a magnetized ICM extracted 
from a high resolution MHD 
cosmological simulation \citep[\cf][and references therein]{mendygraljets}. {The usefulness of such synthetic observations from cosmological simulations is demonstrated, e.g., by \cite{XuRMs}. } {In our case,} the simulated cluster magnetic field  evolved from an initially weak but uniform field  seeded at high redshift (z = 20). Since the cluster and its magnetic field formed dynamically according to the current cosmological paradigm, there are no artificial biases that would influence our analysis other than what result from finite numerical resolution in the magnetic field distribution and omission of ICM radiative or conductive cooling and galaxy feedback.  Those details are largely irrelevant to the exercise at hand, which primarily aims to establish the reliability of estimations for existing ICM magnetic fields using standard RM methods.

We report two complementary sets of analysis experiments. In the first experiments, we obtain and analyze results for a fully sampled background
polarized screen, which offers the optimal information potentially available to an observer. We also analyze a fully sampled screen placed halfway along the line of sight through the cluster (the ``midplane" experiment), in order to confirm assumed scaling relations.  In the second set of experiments, we examine and compare results based on RM measurements for discrete sets of finite-sized, polarized sources, embedded as passive objects in the cluster. We refer to these sources as ``masks'' in our analysis. 

{\ed{ We note that some recent methods \citep{Abell2199,A194} of
estimating intracluster magnetic field properties from observed RMs} use Monte Carlo simulations and a Bayesian analysis with an assumed power law distribution of magnetic field fluctuations.  This approach offers distinct advantages by characterizing some, although not all, of the uncertainties associated with the above issues.  We will discuss these further in the context of our findings in Section \ref{compare}. }

{The outline of the remainder of our paper is as follows:  In \S 2, we describe the physical properties of our simulated cluster and its ICM,  along with beta-law fits for the electron density and magnetic field distributions for use in modeling analysis of the synthetic RM observations. In \S 3 we outline the procedures for translating the RM statistics for a fully sampled background screen into ICM magnetic field  properties. Magnetic field estimates derived from discrete polarized patches (both ``midplane masks'' and ``background masks'') for RM measurements are presented in \S 4.  Further discussion of analysis issues is presented in \S 5, while 
our conclusions are outlined in \S 6.  In the Appendix, we discuss the expected relationships between RM and magnetic field coherence lengths and our methods for estimating these.}

\section{The Test Cluster Properties}

\subsection{ICM Evolution}

{The ICM used in our study is the diffuse baryonic component of the}  $1.5\times 10^{14}~\rm{M}_{\odot}$ cluster \textit{g676} \citep{dolag09a,dolag13} extracted at a redshift of $z \approx 0$ from 
a very high resolution $\Lambda$CDM cosmology simulation ($h = 0.7$, $\Omega_{\mathrm{M}} = 0.3$, and $\Omega_{\Lambda} = 0.7$). The simulation was carried out with an 
MHD implementation of the SPH GADGET-3 code \citep{ClusterPaper}. The magnetic field in that cluster developed from a uniform, primordial field of
strength $10^{-11}$G at z = 20.
For our study here we mapped and centered the evolved cluster ICM and, to allow continued, short-term dynamical evolution mentioned in \S 4, applied a spherical approximation to the total gravitational potential onto a $1~\rm{Mpc}^3$ Cartesian grid (called ``the cluster analysis box'' below)  of uniform resolution with $\Delta x = \Delta y = \Delta z = 1$ kpc.
Details of the \textit{{g676}} cluster and the brief simulation extension that produced the data we used can be found in \cite{mendygraljets} and references therein, {used for the study of jet propagation.}
This cluster was chosen for its lack of recent mergers (the last major merger occurred 7 Gyr prior to the epoch of our RM experiments) and a relatively relaxed morphological appearance, based on synthetic thermal X-ray observations \citep[][]{mendygraljets}. 

\begin{figure} 
   \includegraphics[height=.3\textheight]{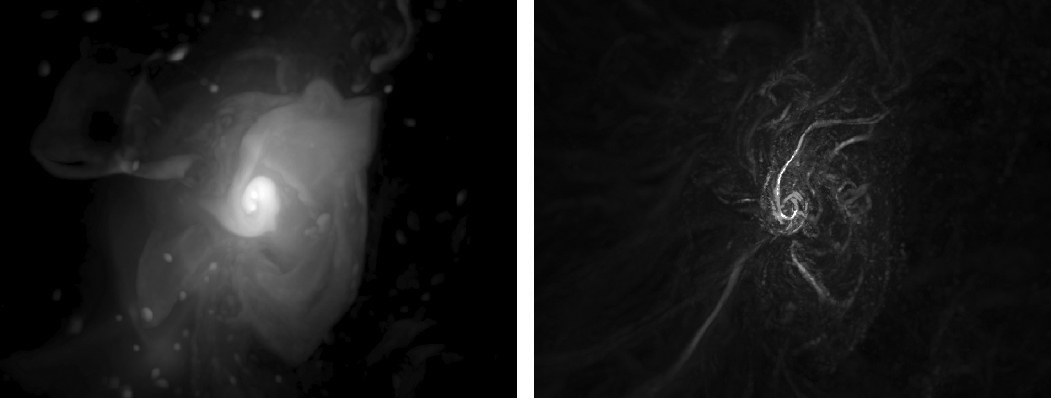}
  
   \caption{Grayscale volume renderings of the {\textit {g676}} ICM along the z-axis showing: ({\bf Left}) log of electron number density ($4\times 10^{-4} {\rm cm}^{-3} \la n_e \la 4\times 10^{-2} {\rm cm}^{-3}$), and ({\bf Right}) log of the magnetic field strength ($0.5\mu{\rm G} \la |B| \la 10\mu{\rm G}$).  
 The volume rendered spans $\sim$1 Mpc. The cluster was extracted from an MHD cosmological simulation (see the text for details). }
\label{fig:densitymap}
\end{figure}

Despite this rather long period devoid of major disruptions and the relatively relaxed morphology, the \textit{{g676}} ICM at $z \approx 0$ still contains significant dynamical
features; in particular, there are large scale ``sloshing" motions  due to gravitational interactions with subhalos, with flow velocities approaching the cluster sound speed ($c_s \approx 650$ km/sec), so Mach$\sim$1.  The
sloshing velocity field is evident in the Figure 5 of  \citet{mendygraljets}. Spiral density and magnetic field structures expected from the sloshing \citep[e.g.,][]{ascasibar06}
are obvious in Figure  \ref{fig:densitymap} of the present paper. Small, isolated density clumps visible in Figure \ref{fig:densitymap} reveal subhalos present in the original cluster formation simulation.
Figure 6 in \citet{mendygraljets} 
provides a synthetic X-ray image of this cluster along the same axis at approximately the same time. Although the cluster is somewhat aspherical in that image, the sloshing structures visible in our Figure \ref{fig:densitymap}  are not obvious.

\subsection{Spherically Modeled Distributions}

While the various deviations from spherical symmetry outlined above are natural and relate to the dynamical state of the cluster, they are not dominant, and for this RM model study we follow the standard practice of constructing symmetric averages for the density and magnetic field distribution. Existing asymmetries will, of course, contribute largely irreducible errors to the RM analyses, although on scales of order 
the cluster core size, both electron density and magnetic field distributions are actually reasonably spherically symmetric and, in the latter case, reasonably isotropic.
Based on spherical averages, the core of the \textit{{g676}} ICM has a radius  $r_{c}\approx 40$ kpc, central mass density, $\rho \sim 10^{-25} ~\rm{g~ cm}^{-3}$ ($n_e \sim 4 \times 10^{-2} {\rm cm}^{-3}$), temperature, $kT \approx~1.6~\rm{keV}$ and pressure, $P_g \approx 2\mathrm{x}10^{-10}\mathrm{dyne~cm}^{-3}$. 

{The spherically averaged ICM mass density at $r = 500$ kpc is $\rho \sim 3\mathrm{x}10^{-28} \rm{g \, cm}^{-3}$, so this radius very roughly corresponds to $R_{500}$. 
The \textit{{g676}}  ICM
is modestly turbulent with core turbulent velocities $\sim 50~\rm{km~s}^{-1}$ \citep{zhurav_11}. These properties correspond to core turbulent pressures $<1$\% of the total pressure.  Outside the core,
turbulent velocities increase to $\sim 100~\rm{km~s}^{-1}$, but still contribute $< 5$\% of the total pressure.
The cluster magnetic field has values locally as large as $12 \mu$G in some strong central filaments visible in Figure \ref{fig:densitymap}, but field strengths more typically fall into a range, $B \sim 0.5 - 4 \mu$G. The RMS core field strength is $\approx 2 \mu$G, corresponding to a magnetic pressure in the cluster core $P_B\sim10^{-13}\mathrm{dyne~cm}^{-3} \sim 0.1\% P_g$. Thus, the central plasma $\beta_p = P_g/P_B \sim 10^{3}$. Field strengths decrease outward, so that $B\sim 0.1\,\mu$G near $r \sim 500$ kpc, producing a magnetic pressure that is, again $\sim 0.1\% P_g$. 
Evaluating Eqn. \ref{eq:RM-def} while assuming $B_{\parallel} \sim B \sim 2 \mu$G and $L \sim 2\times 40$ kpc 
as uniform values through the core would
produce $RM \sim 4000$ rad m$^{-2}$ as a fiducial RM. The analogous  result in the cluster outskirts would be roughly two orders of magnitude smaller.
Disorder in the magnetic field then reduces the observed RM values from these estimates by factors of a few within the core and by much larger factors over the full cluster (Figure \ref{fig:etaBICs}). }

{Spherically averaged ICM characteristics, represented as functions of cluster radius, $r$, and in projection functions of projected radius, $a$, are a standard approach to cluster RM analysis, and underlie our modeling, as well. Of course, this overlooks real ICM structures that limit the comparisons between the standard models and the physical ICM.} As we set up this basic model, we stress that the specific ICM properties of this cluster are not, themselves, central to our {subsequent, analysis-based} conclusions, since the task is to explore uncertainties in analysis outcomes. That is, the underlying questions being explored have to do with {the uncertainties inherent in} the  observational analysis methods to recover the actual cluster properties, whatever they may be.
It is {also} important to note that the model parameters described below are only used to \emph{interpret} the data; the cluster simulation is \emph{not} based on such symmetry assumptions.

\subsubsection{ICM Electron Density}
For analysis modeling purposes we express the electron density radial distribution, $n_e(r)$, in terms of a spherical
beta-law profile \citep{BetaLaw},
\begin{align}
n_{e}(r) = \frac{n_{e,0}}{\left[1 + (\frac{r}{r_{c}})^{2}\right]^{\frac{3}{2} \beta_c}},
\label{eq:beta-den}
\end{align}
where $\mathrm{n_{e,0}}$ is the central electron density, $\mathrm{r_{c}}$ is the ICM  core radius, and $\beta_c$ represents the galaxy- gas velocity dispersion ratio in the cluster,which is nominally assumed spherically symmetric and isothermal. 
We display in the upper left of  Figure \ref{fig:cprops} the spherically averaged profile of $n_{e}(r)$, as well as the
best fit to Eqn. \ref{eq:beta-den}.
The core radius in the fit is $\mathrm{r_{c}} = 41$ kpc, while $\beta_c = 0.75$. Both are consistent with observed properties  of real ICMs (e.g. \citet{ClusterProperties}).

{At this point it is useful to emphasize that observable measures generally correspond to integrals along lines of sight; that is they represent projections. We generally cannot determine 3D quantities directly, but depend on modeling projections. In a spherically symmetric model the relevant positional variable is the projected radius from the cluster center, $a$, rather than the spherical coordinate, $r$. It is necessary, for example, to distinguish between $\sigma_{B}(r)$, an intensive, 3D quantity and $\sigma_{RM}(a)$, a projected quantity. We may, in principle, be successful in determining $\sigma_{RM}(a)$ from observations, but must model the projection to estimate $\sigma_B(r)$.}
In this context and an isothermal ICM approximation, the projected electron density squared, $\Sigma_{n_{e}^{2}}(a)$ 
provides a convenient proxy for the (observable and projected) thermal X-ray surface brightness distribution as a function of $a$. With a beta-law density profile model 
\begin{align}
\Sigma_{n_{e}^{2}}(a) = \frac{C} {\left[1 + (\frac{a}{r_{c}})^{2}\right]^{3 \beta_c + \frac{1}{2}}}\, 
\label{eq:Sigma-fit}
\end{align}
{where  $C = 2\int_0^{\infty} n_e(r)^2 dr = n_{e,0}^2 ~\sqrt{\pi}~\Gamma(3\beta_c - 1/2)/\Gamma(3\beta_c)$ is the square of the electron density integrated along the line of sight through the cluster center.}
The upper right panel in Figure \ref{fig:cprops}  shows radial profiles of the projected, azimuthally averaged electron density-squared distribution along the three principle axes of our grid along with the  beta-law model curve determined by Eqn. \ref{eq:Sigma-fit} using the azimuthally averaged electron profile shown in the upper left panel of the figure.
Deviations in the projected squared density fit to Eqn. \ref{eq:Sigma-fit}, coming especially from the spherical asymmetries noted above, are as large as 15\% near the cluster center, but  less than 5\% at projected distances beyond the core radius. 

Note that in order to derive empirical estimates for the central magnetic field, the values of $n_{e,0}$, $\beta_c$ and $r_c$ must first be derived from X-ray observations (real or synthetic, as described above).

\subsubsection{ICM Magnetic Field}
\begin{figure}  
    \includegraphics[height=.9\textheight]{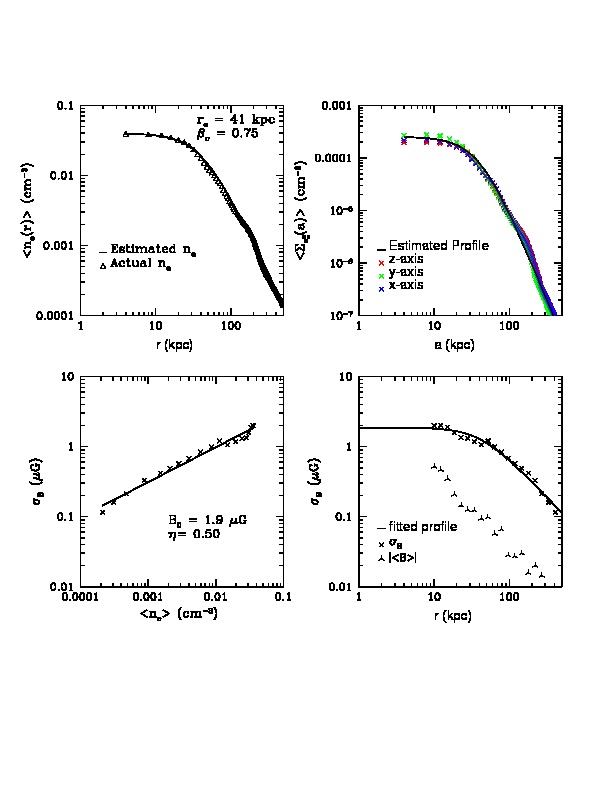}
    \vskip -1in
    \caption{Global properties of the {\textit{ g676}} plasma and magnetic field distributions. ({\bf Upper left}) Spherically averaged electron density, $\langle n_e\rangle$, as a function of  radius, $r$,  in cluster {\textit{g676}} (triangles) with beta-law profile fit by Eqn \ref{eq:beta-den} (solid curve), ({\bf Upper right)} azimuthally averaged projected electron density
squared, $\Sigma_{n_{e}^2}$, as a function of projected distance, $a$, along the analysis grid principal axes (x's) with azimuthal average beta-law profile fit by Eqn \ref{eq:Sigma-fit} overlaid, ({\bf Lower left}) magnetic field strength dispersion, $\sigma_{B}$, versus $\langle n_e\rangle$  along with a fit to $\sigma_{B} \propto \langle n_{e}\rangle^{\eta}$, 
and ({\bf Lower right}) $\sigma_{B}$ as a function of radius (x's)  with the solid line representing a fit using Eqn. \ref{eq:Bprofile}. The mean vector magnetic field magnitude, $|\langle\vec{ B} \rangle|$,is also shown vs radius.  {Note that the bins are logarithmically spaced, so represent larger averaging volumes at larger radii.}}
    \label{fig:cprops}
\end{figure}

We assume any foreground (e.g., galactic) RM has been removed, so that the integral in equation \ref{eq:RM-def} is along the line of sight from the near edge of the cluster to the polarized emission of an embedded source or to the far edge of the cluster for a background source. For the best case scenario, we also assume no other contributions to the RM, e.g., from the immediate environment of the embedded source, or along the post-cluster path to a background source.
Although there are several long magnetic filaments related to ICM sloshing motions in  the \textit{g676} cluster, for modeling purposes, we assume the magnetic field is locally disordered and isotropic. This is a good assumption as long as RMs can be averaged over sufficiently large scales. As seen in the lower right panel of Figure \ref{fig:cprops}, the magnitude of the \emph{mean} vector magnetic field, $|\langle\vec B\rangle|$, is an order of magnitude smaller than the \emph{rms} field strength, $\sqrt{\langle B^2\rangle}$ or its dispersion $\sigma_B$.  In practice, effective isotropy will be valid {\it provided} the RM observations relate to volumes comparable or larger in size than the cluster core or, as it turns out, equivalently, than the correlation length of the magnetic field, as established below.  \emph{However, observations of individual radio galaxies may not cover sufficiently large scales for this assumption to apply;} over smaller scales the $|\langle\vec B\rangle|$ within the observed volume can remain large compared to $\sigma_{B}$, while the observed estimate for $\sigma_{RM}$ will be too small to provider reliable information about the physical dispersion in the local magnetic field strength, $\sigma_B$. 

A different modeling issue arises on scales larger than the core radius.  Since the characteristic strength of the magnetic field (e.g., $\sigma_B$) decreases systematically with distance from the cluster core, as noted in the previous subsection, this behavior must be included for effective modeling. Such a decrease with distance is expected generally in clusters, except where local activity (e.g., AGN jets) may have recently injected a large amount of magnetic flux. 
The cluster-scale variation in the  magnetic field strength has been commonly
expressed in terms of a scaling relation between the magnetic field strength and the ICM density  \citep[e.g.,][]{ryu08,RadMag,Dolag01}. There are several plausible physical arguments
for such a dependence. For instance, flux freezing of a tangled magnetic field during compression would yield $B \propto \ell^{-2} \propto V^{-2/3} \propto n_e^{2/3}$,
 where $V \sim \ell^3$ is the volume containing the magnetic flux. Accounting for work done on a disordered field by the ICM during adiabatic compression would yield $B \propto V^{-1} \propto n_e$, while, alternatively assuming during compression that the magnetic energy maintains a fixed ratio with turbulent energy would lead to $B\propto n_e^{1/2}$. Rather than assume a particular scaling choice, in this paper, we will consider $\sigma_B \propto n_e^\eta$, with $1/2 \la \eta \la 1$. Various MHD cluster formation simulations have shown results roughly consistent with such scalings \citep[e.g.,][]{sphmag,vazza17b}. 
 

In general, RM observers would not know $\eta$ {\it a priori} and would, therefore, have to estimate $\eta$ based on theoretical considerations or simulation results.  In cases where there is extensive RM data, one can attempt to estimate $\eta$ by comparing the radial dependence of $\sigma_{RM}$ to the radial dependence of $n_e$ \citep[e.g.,][]{Dolag01,guidetti08,Bonafede10,A194}.  However, the ability to do this depends on the unknown distribution of polarized sources along the line of sight, the assumption of an unchanging coherence scale, and the lack of any local effects around the embedded radio galaxies.  For {\textit{g676}}, since we have full knowledge of $\vec{B}$,  we can actually determine an approximate value of $\eta$ as follows. First, we
calculate $\langle n_{e}(r) \rangle$ and $\sigma_{B}(r)$ within spherical shells of fixed thickness and uniform logarithmic spacing in $r$,  then compute a least squares fit for log $\sigma_{B}(r)$ vs log $\langle n_{e}(r) \rangle$ within the cluster. The result, shown in the lower left panel of Figure \ref{fig:cprops}, produces  $\eta = 0.5$, with a central, core magnetic field strength dispersion, $\sigma_{B,0} = 1.9 ~\mu$G. 

\begin{figure}  
\begin{center}
   \includegraphics[height=.6\textheight]{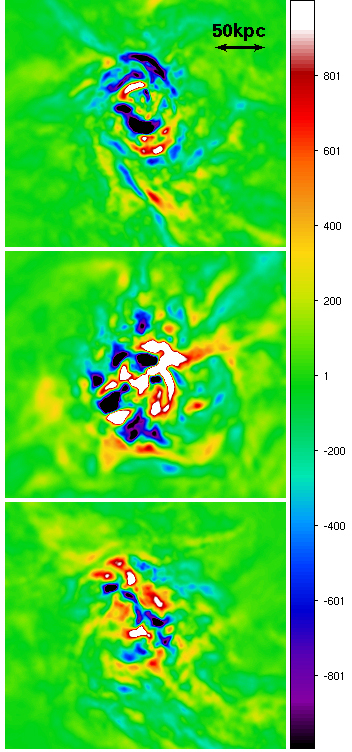}
   \end{center}
   \caption{RM map of the projected central 200 kpc of the {\textit{ g676}} ICM obtained by integrating through the whole extracted analysis box ( 1 Mpc length) along the z-axis (top), y-axis (middle), and x-axis (bottom). 
The colorbar unit is $\mathrm{rad~ m^{-2}}$.} 
   \label{fig:rmmap}
\end{figure}

Using  such a density scaling, along with the beta-law density profile model in Eqn. \ref{eq:Sigma-fit} we can express the spherically averaged magnetic field strength  dispersion as a function of radius, $r$, by
\begin{align}
\sigma_{B}(r) = \frac{\sigma_{B,0}}{(1 + (\frac{r}{r_{c}})^2)^{\frac{3}{2}\eta\beta_c}},
\label{eq:Bprofile}
\end{align}
where $\beta_c$ and $r_{c}$ are obtained from the electron density distribution as described in the  previous subsection.  
The radial profile of $\sigma_B$ for {\textit{g676}} is shown in the lower right panel of Figure \ref{fig:cprops} along with a least squares fit
to Eqn. \ref{eq:Bprofile} allowing both $\eta$ and $\sigma_{B,0}$ to vary. As expected from the excellent fits
for the beta-law density form and the previously established log $\sigma_B$ vs log $n_e$ form, the best fit parameters in {\textit{g676} are consistent again with $\sigma_{B,0} = 1.9~ \mu G$ and $\eta = 0.5$. 

\section{Estimating the Magnetic Field Distribution from the RM Distribution}
\subsection{The Basic Model}

The previous section outlined the actual 3D magnetic field and electron density distributions in our simulated test cluster, {\textit {g676}}, and described beta-law model, spherically symmetric radial fits for those distributions. The observational challenge is to recover the (spherically averaged) magnetic field properties as a function of radius, $r$, in a cluster from an observed distribution of RM, which is a 2D, projected quantity. In particular, the objective is to obtain estimates for $\sigma_B(r)$ and then $\sigma_{B,0}$ from the RM dispersion, $\sigma_{RM}(a)$, as a function of projected radius, $a$. If the magnetic field is disordered and isotropic, \S 1 showed that this problem reduces to obtaining reliable measures for the RM dispersion, $\sigma_{RM}(a)$ along with an estimate of the magnetic field coherence length, $\Lambda(r)$, within the cluster, and assuming some value or range for $\eta$. {We emphasize that, in practice one cannot directly determine $\Lambda(r)$, which is a property of the 3D magnetic field, but must rely on estimating the {\it{projected}} RM coherence length, $\Lambda(a)$ and modeling a connection between these lengths. Generally, if the magnetic field is isotropic, it is assumed that $\Lambda(a) \approx \Lambda(r) = (3/4)L_{int}(r=a)$. We will follow that convention.} 

{To summarize, the procedure we used in this work to estimate from observations the rms value of the magnetic field strength at the cluster center,} $\sigma_{B,0}$, is to:
\begin{itemize}
\itemsep -5pt
\item{Determine the central density $n_{e,0}$, and $\beta_c$ and $r_c$ from X-ray observations;}
\item{Estimate $\sigma_{RM}(a)$ and fit it to a beta-law model in order to estimate the central $\sigma_{RM,0}$;}
\item{Determine an estimated value for the RM coherence length, $\Lambda$, and make an assumption about whether it is a constant or a function of cluster radius;}
\item{Assume a value for the magnetic field density scaling parameter, $\eta$}
\item{Derive from these inputs $\sigma_{B,0}$, as outlined below.}
\end{itemize}


 To estimate the {theoretical} central RM dispersion, $\sigma_{RM,0}$, {for our cluster} we average RM and RM$^2$ azimuthally  with respect to the cluster center as functions of projected cluster radius, $a$, and calculate $\sigma_{RM}(a) = \sqrt{\langle RM^{2} \rangle|_a - \langle RM \rangle^{2}|_a}$. 
Assuming an isotropic magnetic field along with beta-law radial electron density and magnetic field models, we integrate through the full cluster to obtain from Eqn. \ref{eq:Bprofile} the actual, theoretical radial RM dispersion distribution.  { Following \citet{Dolag01},} with a fixed $\Lambda = \Lambda_0$ within the cluster and observational path lengths through the cluster, $\ell \gg \Lambda_0$, this would give
\begin{align}
\sigma_{RM}(a) = \frac{\sigma_{RM,0}}{\left[1 + (\frac{a}{r_{c}})^{2}\right]^{\alpha_1}},
\label{eq:RM-model1}
\end{align}
where the exponent, $\alpha_1 =\frac{3}{2}(1+\eta)\beta_c-\frac{1}{4}$ and $\sigma_{RM,0}$  is the central RM dispersion.  Note, also, that  {for equation \ref{eq:RM-model1} to be consistent with $\sigma_{RM}$ values determined from observations that will be used in practice to estimate $\sigma_B$}, the observational distribution of RM sight lines must span projected scales significantly exceeding $\Lambda_0$.  

Figure \ref{fig:rmmap} displays the RM distributions in {\it{g676}} (at 1 kpc resolution) obtained by using a polarized screen with a 100\% covering factor behind the cluster, viewed alternately along the three analysis grid axes.  We can now use the central RM dispersion, $\sigma_{RM,0}$, in Eqn. \ref{eq:RM-model1} to derive the central magnetic field dispersion, $\sigma_{B,0}$, in Eqn. \ref{eq:Bprofile} for constant $\Lambda = \Lambda_0$ by the relation 
\begin{align}
\sigma_{B,0} =  \sigma_{RM,0}\, \frac{\sqrt{3}}{812 \, \pi^{1/4}} \, n_{0}^{-1} \, \Lambda_{0}^{-1/2} \, r_{c}^{-1/2} \, 
\sqrt{\frac{\Gamma(\alpha_1+1/2)}{\Gamma(\alpha_1)}}.
\label{eq:sigmaRM0}
\end{align}
 The rms variation in magnetic field as a function of radius $\sigma_{B,r}$ then follows from Eqn. \ref{eq:Bprofile}.   

We point out in the Appendix that $L_{int}(r)$ (so $\Lambda(a)$) actually increases with radius in our test cluster, or as a conveniently simple,  alternate expression, decreases with spherically averaged, mean density (or for $\Lambda(a)$, projected mean density).  If we adopt the scaling relation
suggested  in \S 2; namely, $L_{int} \propto n_e^{-\eta/2}$, we can model the projected RM coherence scale, $\Lambda(a)$, as 
\begin{align}
\Lambda(a)  = \Lambda_0 \left[ 1 + (\frac{a}{r_{c}})^{2}\right]^{\frac{3}{4}\beta_{c}\eta},
\label{eq:varlamb}
\end{align}
where $\Lambda_0$, now refers to $\Lambda(a=0)$.
%
Eqn. \ref{eq:RM-model1} then becomes slightly modified to
\begin{align}
\sigma_{RM}(a) = \frac{\sigma_{RM,0}}{\left[1 + (\frac{a}{r_{c}})^{2}\right]^{\alpha_2}},
\label{eq:RM-model2}
\end{align}
where $\alpha_2 = \frac{3}{2}(1+\frac{3}{4}\eta)\beta_c-\frac{1}{4} = \alpha_1 - \frac{3}{8}\eta\beta_c$. For characteristic parameters $\beta_c \sim 3/4$, $\eta \sim 1/2$, the difference between $\alpha_2$ and $\alpha_1$ is roughly 10\%. $\sigma_{RM,0}$ can still be obtained using Eqn. \ref{eq:sigmaRM0}, provided $\Lambda_0$ is found from Eqn.~ \ref{eq:varlamb} and $\alpha_2$ replaces $\alpha_1$ in the $\Gamma$ function arguments. The latter substitution leads to a small renormalization of the RHS of Eqn. \ref{eq:sigmaRM0} ($< 3$\% for the above characteristic $\beta_c$, $\eta$). On the other hand, we note that in the case of \textit{g676}, that $\Lambda(a)$ varies by a factor of 3 between $a = r_c$ and $a = 10r_c$, so including the variation of $\Lambda$ with radius in fitting for $\sigma_{RM,0}$ can significantly change the estimates of $\sigma_{B,0}$.  
%
%

We emphasize again that the above relationships assume both that the magnetic field itself is reasonably isotropic, so that $(|\langle \vec{B}\rangle|)^2 \ll \langle B^2\rangle$, and that the RM values used in computing $\sigma_{RM}$ broadly sample independent parts of the RM distribution. If the RM distribution is not well sampled on large enough scales ($\ga \Lambda$), then even for an isotropic magnetic field, the measured properties will lead to $(\langle RM\rangle)^2 \sim \langle RM^2\rangle$.  {In particular, it is critical to include lags, $|\vec{\Delta a}|$, between sampling points that satisfy $|\vec{\Delta a}| \ga \Lambda$.}  {\citet{obsmodel1} show the dependence of $\langle RM\rangle / \sigma_{RM}$ as a function of the utilized sampling space.}  In the case where the sampling region is too small, the estimated $\sigma_{RM}$ (= $\sqrt{\langle RM^2\rangle - \langle RM\rangle^2}$ will generally be reduced, so will lead to underestimates for $\sigma_{B,0}$ through Eqn. \ref{eq:sigmaRM0}.

As noted above, both $\beta_c$ and $r_{c}$ in such an analysis are established observationally from the X-ray surface brightness distribution. From synthetic X-ray observations of our simulated cluster we found in \S 2.2 values $\beta_c = 0.75$ and $r_c = 41$ kpc. The remaining parameters in our RM models are $\sigma_{RM,0}$, $\eta$ and $\Lambda_0$. We defer discussion of estimates for
$\Lambda_0$ to the following subsection. The central RM dispersion, $\sigma_{RM,0}$ is obtained from observations by fitting
 an empirical RM
distribution to Eqn. \ref{eq:RM-model1} (or alternatively Eqn. \ref{eq:RM-model2}, if $\Lambda$ is a function of radius). 

The $\eta$ parameter needed in Eqn.\ref{eq:BRMSeval} comes from interpreting the fitted slope as $\alpha_1$ for a fixed RM coherence length, $\Lambda(a) = \Lambda_0$, or as $\alpha_2$, if the RM coherence length follows Eqn. \ref{eq:varlamb}. For reference we recall here that in \S 2 we obtained $\eta = 0.5$ from the 3D density and magnetic field distributions in our cluster. Using $\beta_c = 3/4$ and $\eta = 1/2$ as nominal parameters, the exponents in Eqn.\ref{eq:RM-model1}, Eqn. \ref{eq:varlamb} and Eqn. \ref{eq:RM-model2} become $\alpha_1 = 23/16\approx 1.44,~\frac{3}{4}\beta_c\eta = 9/32\approx 0.28~\rm{and}~\alpha_2= 83/64\approx 1.30$ respectively. Outside the cluster core, where $a/r_c \gg 1$ the theoretical RM dispersion would scale with projected radius as $\sigma_{RM} \propto (a/r_c)^{-2.9}$ for constant $\Lambda$. Including the previously outlined density scaling for $\Lambda$ would lead to $\Lambda(a)\propto (a/r_c)^{0.56}$ and $\sigma_{RM}(a)\propto (a/r_c)^{-2.6}$. These are rather strong radial scalings, especially for the RM dispersion. One obvious consequence that we address in the next subsection
is that measurements depending on $\sigma_{RM}$ that include the cluster core will be dominated by contributions from the core.  

In preparation for discussion below, we also point out the sensitivity of solutions for $\sigma_{B,0}$ to the $\eta$ parameter. That sensitivity comes through its presence in the RM distribution shape parameters, $\alpha_{1,2}$ in Eqns. \ref{eq:RM-model1} and \ref{eq:RM-model2}. In particular, as $\eta$ decreases, both $\alpha_{1,2}$ decrease, so the radial variation in $\sigma_{RM}$ is reduced. Thus, given values of $\sigma_{RM}$ at finite radii, $a$, smaller $\eta$  lead to smaller values for $\sigma_{RM,0}$ and thus smaller values for $\sigma_{B,0}$.

Finally, for convenience, we rewrite Eqn. \ref{eq:sigmaRM0} in terms of 
approximate values for the {\textit {g676}} cluster. In particular,
\begin{align}
\sigma_{B,0}~\approx~2~\mu{\rm Gauss} ~\frac{\sigma_{RM,0}/(1000~ {\rm rad~m}^{-2})}{(n_0/0.04~{\rm cm}^{-3})(\Lambda_0/17~{\rm kpc})^{1/2}(r_c/41~{\rm kpc})^{1/2}} .
\label{eq:BRMSeval}
\end{align}
All our analysis fits below assume a central electron density, $n_0 = 4\times 10^{-2}$ cm$^{-3}$ and cluster core radius, $r_c = 41$ kpc.

\section{RM Experiments Using Discrete Regions} 
In the previous section, to obtain a ``theoretical'' distribution for $\sigma_{RM}$ we assumed that data were available from a well-sampled background polarized screen spanning the entire cluster.  This is not achievable in practice, so we conducted a series of experiments restricting those conditions in ways that mimic common experiences.  In each case, we generated a finite set of discrete background regions (patches), each with their own values of projected radius, $a_i \doteq (1/A_i) \int a~ dA_i$  with $A_i$ the area of an individual patch.  For each patch, there was an associated, ``measured''  $\langle RM_i \rangle$ and $\sigma_{RM,i}$.


In the next section, \S \ref{bscreens}, we will look at fully sampled background screens, using discrete regions in the form of annuli surrounding the cluster center.  Although the cluster is  then still fully sampled, the discretization of the RM information can introduce problems, as discussed below.  In \S \ref{masks} we will use rectangular masks of various sizes placed at different locations along the line of sight, to approximate RM observations of individual radio galaxies embedded in the cluster.
 
 {In order to reduce ``pixelation'' issues
associated with the 1 kpc$^3$ discrete voxel size of the MHD simulation that produced the ICM being modeled, all subsequent 2D images presented below are averaged over 2$\times$2 pixels, for an effective observing resolution of 2~kpc.} This is still sufficient to sample the RM structure, but does impose a fine scale cutoff, for example, to a structure function analysis we carried out on synthetic RM data.


\subsection{Discrete Background Screens}
\label{bscreens}

Here we assume the observer can sample RMs on discrete background screens. Rather than the full sampling available in our ``theoretical'' RM scenario in \S 3, discrete patches do not necessarily allow full sampling of the RM distribution. The coverage depends on how the patches/screens are constructed.  Information can be lost that may or may not influence the estimation of $\sigma_{RM,0}$.  Most important in this 
is the maximum length of the available lag vectors, $|\Delta\vec{a}|$,
since without large lags, $|\Delta\vec{a}| \ga \Lambda$, all the RM values within a patch will be correlated.  In that case, $\langle RM\rangle^2 \sim \langle RM^2\rangle$ within each patch,  {even if the underlying magnetic field is disordered and isotropic on scales beyond $L_{int} \approx \Lambda$.} 
Then $\sigma_{RM}$ will be reduced from its true, physical value needed to estimate $\sigma_B$ properly.

 For these experiments we first create  {ideal, background} patches in the form of annuli around the cluster center,  uniformly spaced in $\log{a_i}$ with the minimum $a_i$ placed somewhat inside the projected core radius. 
 We explored the consequences of varying ring thicknesses, $\delta a$, and found converged results so long as $\delta a \ge 10$ kpc. 
Thus, we limit our discussion to the illustrative $\delta a = 10$ kpc case. Note that due to the logarithmic spacing, there is some overlap in the annuli at the smallest radii, and gaps between the rings at the largest radii. 

The left two panels of Figure \ref{fig:etaBICs} summarize results of the $\delta a = 10$ kpc annular rings viewed along the three primary grid axes.
The top panel shows measured
$\sigma_{RM}(a_i)$ values plus fits to the form in 
Eqn. \ref{eq:RM-model2} for variable $\Lambda$ allowing $\eta$ as a free parameter (see Table \ref{tab:BICs} for fitting summaries).
 
We checked whether the annuli have adequately sampled the largest scales of the RM fluctuations by examining the ratio $|\langle RM\rangle|/\sigma_{RM}$,  {which, in reality, should be small}.  The computed values for each annulus are shown in the bottom-right panel of Figure \ref{fig:etaBICs}.
The median  $|\langle RM\rangle|/\sigma_{RM} \sim 0.2$. Even with the apparent scatter, the values are small enough that the associated $\sigma_{RM}$ still represent  {reasonably appropriate} measures for an isotropically disordered magnetic field behavior.

Table \ref{tab:BICs} provides an analysis summary from the annular background screen experiment. Results are given both for constant $\Lambda = \Lambda_0$ and radially varying ($\Lambda = \Lambda(a)$). In each  case fits are shown with { fixed (preset) value of the magnetic field density scaling parameter}, $\eta = 0.5$ and also with $\eta$ as a free, fitting parameter. 
These estimates for $\Lambda_0$ using annular background screens are consistent with what we found from the full background screen in the previous section, for the same assumptions. 
The results in Table \ref{tab:BICs} show that if we fix $\eta$, then we get a range of only 10-20\% in the values derived for $\sigma_{B,0}$ from the various projections.  ( {Note, however, that the average for $\sigma_{B,0}$} in the fixed $\Lambda$ case is off by a factor of 1.6).  We see further that if $\eta$ is allowed to vary in the fitting, estimates of $\sigma_{B,0}$ will span a range $\approx$ 3.  \textit{Thus, with
full RM coverage and independent knowledge of $\eta$ it is possible to derive estimates of the magnetic field strength only within a range $\approx$ 3.}

The strong relationship that exists between the derived
$\sigma_{B,0}$ and $\eta$ from these solutions is illustrated in Figure \ref{fig:etaBIC2s}. 
This arises from the sensitivity of $\sigma_{RM}(a)$ to $\eta$ through the $\alpha_{1,2}$ shape parameter in Eqn. \ref{eq:RM-model1} or Eqn. \ref{eq:RM-model2}. Smaller $\eta$ leads to smaller $\alpha_{1,2}$, which makes the form of $\sigma_{RM}(a)$ ``stiffer''. Thus, for example,
a fit to $\sigma_{RM}(a)$ in  Eqn. \ref{eq:RM-model2} using $\eta = 0.5$ increases between $a/r_c = 5$ and $a/r_c = 0$  by a factor 4 times larger than it does using $\eta = 0$. 
An additional contribution to the $\eta$ dependence of $\sigma_{B,0}$ in Eqn. \ref{eq:BRMSeval}, when $\Lambda$ varies with cluster radius, comes from the $\eta$ dependence of $\Lambda(a)$ itself in Eqn, \ref{eq:varlamb}.  {This
important degeneracy between the magnetic field model parameters has been extensively discussed in the literature \citep[e.g.,][]{obsmodel1,guidetti08,Bonafede10,Abell2199}.} In general,  higher $\sigma_{B0}$ correlate with higher $\eta.$}

Thus, even estimates using near perfect RM coverage are \textit{not} reliable to better than a range of 3 unless $\eta$ is assumed {\it{correctly}} within an unmeasurable range of values.  Looking more closely at the variable-$\Lambda$ fits in Figure \ref{fig:etaBIC2s}, we see that there is little change from the derived value of $\sigma_{B,0}$ compared to the fixed $\Lambda$ case, although $\eta$ increases by $\sim$0.1 - 0.2 .  If $\eta$ were to be held fixed  {in the fits}, however, as we suggest below, then the assumption of a variable $\Lambda(a)$ can result in a factor of up to 2 decrease in $\sigma_{B,0}$.

We performed an additional test of the robustness of the annular sampling procedure by placing the polarized screens at the midplane, instead of behind the cluster.  
We again viewed the screens along the three principal grid axes from both directions. When the computed $\sigma_{RM,0}$ values for each mid-plane (MP) experiment were renormalized by a factor $\sqrt{2}$ to adjust for the shorter path (factor 1/2) to the mid-plane, $\sigma_{B,0}$ estimates were consistent with
those for the above annular background screens (BG) within the statistical uncertainties. This good match was possible because paths through half the cluster still incorporate multiple RM coherent lengths, and, just as important, we know the actual path to the screens. The rescaled mid-plane annular ring screen partition $\sigma_{B,0}$ values are shown in Figure \ref{fig:etaBIC2s} with ``MP'' (mid-plane) identifications.  

To examine whether these results were dependent on the specific use of annuli, we repeated the experiment by breaking up the background screen into a set of square patches.  As long as the square patches were larger than $\Lambda_0$ ($\sim 20$ kpc), the average derived values of $\sigma_{B,0}$ were consistent with the actual values, although with larger scatter between the different projections than the annular results, which spanned scales $>> \Lambda_0$.  When the square patches were smaller than $\Lambda_0$, the derived $\sigma_{B,0}$ values were consistently too small (a factor of 2 for 12~kpc boxes), consistent with expectations.  As noted earlier, the value of $|\langle RM\rangle_i|/\sigma_{RM,i} \geq 1$ provided a good indicator that the larger scales of RM variations were not being adequately sampled by the 12~kpc boxes, since they are smaller than the actual magnetic field coherence length.

\subsection{Embedded Masks}
\label{masks}

To approximate the type of information available from RM  observations of  {cluster-embedded} radio galaxies, we conducted a series of experiments using rectangular ``masks'' that provide a sparse, but observationally realistic sampling of the cluster RM distribution.  The first type of experiment simply sampled the RM distribution of rectangular masks placed randomly along and at right angles to the line of sight at different projected positions within the cluster.  The second type of experiment looked at the RM distribution along the lines of sight to a central radio galaxy that was actually evolved by MHD simulation within this ICM. We defer to a subsequent study the RM consequences from physical displacement of adjacent ICM. Here we simply insert a 2D mask into the undisturbed ICM that matches the silhouette of the radio galaxy. 

Our analysis of these embedded masks was similar to those discussed above. Statistics from synthetic RM observations were computed across individual masks to determine values for $\sigma_{RM}(a_i)$, with $a_i$ the mean projected  cluster radius of each mask. 
The ensemble $\sigma_{RM}(a)$ distributions from the masks were then fit to Eqn. \ref{eq:RM-model2} to find $\sigma_{RM,0}$. They were  also incorporated into a 2$^{nd}$ order RM  structure function (Eqn. \ref{eq:RM-sf} in the Appendix) to find $\Lambda(a)$.
Then $\Lambda(a)$ as represented in  Eqn. \ref{eq:varlamb} was used to estimate $\Lambda_0$. $\sigma_{RM}(a_i)$ errors were assigned for the purposes of finding $\sigma_{RM,0}$, including a statistical component $\sim \frac{1}{\sqrt{N_{dof}}}$, where $N_{dof}$ represents the number of independent
observing beams across a mask  plus a constant $5~ {\rm{rad ~m^{-2}}}$ component 
representing the uncertainty in the measurement of $\sigma_{RM}$. Although somewhat arbitrary, most observations would have uncertainties at least this large These two error components were added in quadrature to get the total error used. Finally, Eqn. \ref{eq:BRMSeval}, renormalized by a factor $\sqrt{2}$ to account for the assumed midplane location, was used to translate $\sigma_{RM,0}$ and $\Lambda_0$ into $\sigma_{B,0}$.

\subsubsection{Embedded Passive Masks}
\label{passive}

In this experiment,  we randomly distributed  multiple polarized, rectangular planar masks within our cluster, with the mask normals aligned to the line of sight. Individual, rectangular masks had variable aspect ratios. Their side lengths ranged between 5 kpc and 50 kpc  representing the extent of typical RM maps of cluster galaxies, e.g. \cite{Abell2199}. 
Their average extent, $\approx 20$ kpc, was comparable to the RM coherence length in the cluster core, $\Lambda_0 \approx 17$ kpc, but smaller than the cluster core radius, $r_c \approx 40$ kpc. 

The randomly distributed, rectangular mask experiments involved ensembles of 3, 8 and 14 masks.  Figure \ref{fig:randsource} illustrates an example synthetic RM observation from the 8-mask experiment as they appear along the z axis.  RM distributions  were obtained for each mask ensemble  
projected along all three grid principal axes from both directions (so a total of six views). To derive $\sigma_{RM,0}$, we assumed the masks to be in the cluster mid-plane, although, in fact, they existed at random displacements with respect to that plane. Averaged over all views, the sources are centered around the midplane.

Figure \ref{fig:sigfit} presents the $\sigma_{RM}$ statistics for the 8 random, embedded mask  RM distribution shown in Figure \ref{fig:randsource} along with curves representing two fits of Eqn.\ref{eq:RM-model2} to the data. The solid curve includes $\eta$ as a fitting parameter, so that the shape parameter, $\alpha_{2}$, of $\sigma_{RM}(a)$ is part of the fit.  It is obvious in this case that the available data are simply inadequate to obtain a meaningful value for $\eta$ as part of the fitting effort. The best fit value, $\eta = - 0.37$, is unphysical, while the accompanying value for $\sigma_{RM,0} \approx 152$  rad/m$^2$ is only about $\frac{1}{5}$ the values $\sim 700$  rad/m${^2}$, obtained from the background screen.  
The dashed curve in Figure \ref{fig:sigfit} represents a fit to the same random mask RM data, but with fixed $\eta=0.5$, matching the physically determined value in this cluster. 
The associated estimate for $\sigma_{RM,0} = 400~{\it rad/m}^2$ is an improvement, but still only about 60\% the expected value.

 We restrict our remaining analysis of the random mask experiments to the physically established $\eta = 0.5$ (along with  a radially dependent $\Lambda(a)$ represented in Eqn. \ref{eq:varlamb}). 
Table \ref{tab:passsource} summarizes those experiments.  It lists the $\langle\Lambda_0\rangle$ and $\langle\sigma_{RM,0}\rangle$ values obtained from six combined views of each embedded source ensemble and associated, representative estimates for $\langle\sigma_{B,0}\rangle$ along with ranges of these values
for different views after trimming the two most-extreme values from the calculation. This provides a very conservative estimate of the uncertainties.
Several points are clear. All of the experiments produce $\sigma_{RM}$ estimates (and consequently $\sigma_{B,0}$ estimates) that are both highly uncertain and significantly reduced from correct values for this ICM. Similar to our full screen experiments with square box partitions (\S 4.1), this results from the fact that these masks are too small to sample independent portions of the RM distribution adequately. Accordingly, the ratio, $|\langle RM\rangle|/\langle\sigma_{RM}\rangle$, is simply too large to allow reliable translation of $\sigma_{RM}$ into $\sigma_B$ through Eqn. \ref{eq:BRMSeval}.

The results in Table \ref{tab:passsource} from the 8 source experiments are no better than the 3 source experiments.  This demonstrates that the undersampling within each patch of the larger scale fluctuations, rather than their limited number is the critical limitation.  Additional scatter is also introduced in this experiment because the masks are not actually at the assumed fixed, midplane location.

\subsubsection{Embedded Central AGN Jet-formed Cavity Masks}
\label{active}
In another set of experiments we measured the foreground RMs in front of the cavities produced in numerical simulations where intermittent bipolar AGN jets were injected at the center of the cluster (run \textit{g676} in \cite{mendygraljets}).  The use of intermittent jets, with a 50\% duty cycle, resulted in distinct, ``fat" cavities with axial ratios of $\sim 2:1$.  In the experiments we describe here, we used those AGN simulations only to define the silhouette of the two cavities;  the ICM for our purposes was exactly the same as in our previous experiments, with no central AGN.  The cavity silhouettes thus define our ``masks" within which to measure RMs.  They evolved in time, with a greatest extent of approximately 120~kpc on each side. At each of two times, 79~Myr and 92~Myr, we used the corresponding ICM, evolved without the presence of the radio galaxy; in this way our results reflect only the time dependent behavior of the ICM, without the influence of the radio galaxy. Analysis of the RM structure with the actual jet-modified ICM will be discussed in a future work. The orientation of the jet axis was arbitrarily set to $\approx 45^o$ from the z-axis of the analysis grid, and the mask planes included the major axis of each cavity and the joint normal to that axis and the line of sight.
  ~We observed the bipolar masks bidirectionally along all three analysis grid axes, and, in addition, from two arbitrary directions normal to the jet axis (so 8 total views) at each time. Figure \ref{fig:rmjetmap} shows RM distribution maps of the  masks viewed along the z axis at the two times mentioned above.  

For the RM analysis, we partitioned each projected mask into annular sectors centered on the position of the AGN at the cluster center. 
We present  the results utilizing the radially dependent $\Lambda(a)$ in Eqn. \ref{eq:varlamb} with $\eta = 0.5$.  In the translation of $\sigma_{RM,0}$ to $\sigma_{B,0}$ both sides of the AGN structure were assumed to be in the mid-plane of the cluster, so that a renormalization factor $\sqrt{2}$ was applied to Eqn. \ref{eq:BRMSeval}.  In actuality, for views down the grid axes, the distance along the line-of-sight varied across the mask, but in a way which the average line-of-sight is approximately equivalent to the midplane line-of-sight.  Those details turn out not to be particularly important, compared to the differences in RMs from the different viewing angles. 

Table \ref{tab:agnmasks} lists the values for $\langle\sigma_{B,0}\rangle$ averaged over all eight views, as well as the range in $\sigma_{B,0}/\langle\sigma_{B,0}\rangle$ for the individual views.  Once again the two extreme  values of $\sigma_{B,0}$ were excluded from each range. At both observation times, the mean value, $\langle\sigma_{B,0}\rangle$, comes within roughly 10\% of the associated physical $\sigma_{B,0}$ of the cluster. This is a consequence of the masks covering spatial scales significantly larger then $\Lambda_0$. Although the averages are quite accurate, the individual derived values $\sigma_{B,0}$ span more than a factor of two. This reflects the fact that the ICM is not, in reality, statistically homogeneous, even on scales larger than $\Lambda_0$, with distinct features such as the magnetic filaments evident in Figure \ref{fig:densitymap}. Different views therefore can yield quite different estimates of $\sigma_{B,0}$.

\section{Discussion}

The most important finding from these studies is that the strength of the central magnetic field can only be determined to within a range $\sim$ 3 even in the ideal, practically unrealistic case of a fully sampled background of rotation measures.   In any more realistic situations, with very partial sampling, estimates of the central field strength are much more uncertain, ranging up to a factor of $30$ in our experiments.  The critical factors leading to these uncertainties are:  


\underline{a) ICM inhomogeneity}. The richness of magnetic field structures, even in clusters that have not experienced a recent significant merger, is evident in Figure \ref{fig:densitymap}.  In an ideal world, these real inhomogeneities would become statistically inconsequential on large enough scales.  In practice, clusters will have inhomogeneities that span scales as large as the cluster itself due to the ongoing intermittent accretion along filaments, etc.  We studied the influence of these inhomogeneities on $\sigma_{B,0}$ determinations by viewing the cluster from different directions under a range of circumstances. These included the ideal of a fully-sampled polarized background screen spanning the cluster, and the more realistic case of sparse sampling.  When averaging over multiple, individual experiments in a class, an accurate value of $\sigma_{B,0}$ can sometimes be recovered (e.g., 10-20\%), although this commonly involves the unrealistic assumption that the physical parameters that characterize the distribution, $\eta$ and $\Lambda_0$, are well-determined. Even then, any individual experiment will typically yield a value within a range of no less than a factor $\approx$ 3.

\underline{b) Magnetic field, density scaling}. The derived $\sigma_{B,0}$ is sensitive to the magnetic field density scaling parameter, $\eta$ ($B\propto n_e^{\eta}$).  While MHD  cluster formation simulations show that there can be a characteristic scaling, $\sigma_B \propto n_e^\eta$, in practice, $\eta$ cannot be determined accurately from observations.   Fitting for $\eta$ depends on subtle shape differences in the profile of $\sigma_{RM}(a)$, as described in Equations \ref{eq:RM-model1} and \ref{eq:RM-model2}.  Even for fully sampled background screens, allowing $\eta$ to vary in fitting leads to a large range in derived $\sigma_{B,0}$ as seen in Figure \ref{fig:etaBIC2s}.  The problem is greatly exacerbated for sparse sampling similar to that available from actual observations.  In that case,  the shape of $\sigma_{RM}(a)$ is poorly constrained, and the unknown position along the line of sight of a source (see point {\bf{\underline{d}}}) adds a large, further complication.  Since many simulations indicate that $\eta \approx 0.5$, our experiments suggest that assuming that value, or something similar, may yield the most accurate estimates of $\sigma_{B,0}$. 

\underline{c) RM coherence length}.  Using Eqns. \ref{eq:sigmaRM} and \ref{eq:BRMSeval}, the magnetic field estimates scale as the assumed value of $\sqrt{\Lambda}$. We estimated $\Lambda$ from the RM structure function, identifying the scale, $\Lambda_{SF}$, at which the slope of the structure function first reached zero from small scales as $\Lambda$.   In the ideal case of the mid-plane screen experiments, we found a range of $\sim$25\% in $\Lambda$ which, therefore, makes only a minor contribution to the error budget.  Randomly placed embedded source experiments have more limited sampling, and the range of derived $\Lambda_{SF}$ and estimated central cluster value $\Lambda_{0}$ is 10-30 kpc. But, since $\sigma_{B,0}$ depends only on $\sqrt{\Lambda}$, this has relatively modest impact compared to $\bf{\underline{a}}$ or $\bf{\underline{b}}$. 
 We note that while this procedure was successful in the current experiment, one can envisage other magnetic field configurations, such as very flat magnetic power spectra, where only the minimum scale is accessible, and a different way of utilizing the RM structure function would be needed \ed{\citep[e.g.,][]{laing08}. }

\underline{d) Line of sight uncertainties}.  Since the positions of individual, embedded sources along the line of sight are unknown and unknowable, (except for cluster center sources) the application of Eq. \ref{eq:sigmaRM} must be based on some assumption.  In our derivations, we assumed that all RMs were integrated through the entire cluster, or for our midplane experiments, half the cluster.  For a large enough sample of sources with RMs, the midplane would be a good approximation.  However, as each individual source contributes to the measurements $\eta$ and $\Lambda(a)$, the assumption of a midplane location can lead to large uncertainties and even apparent non-physical behaviors (e.g., a derived negative fit for $\eta$).  In the best case, where we fix $\eta$ and the $\Lambda(a)$ behavior, Table \ref{tab:passsource} shows that a range of 3 in the fit for the trimmed $\sigma_{B,0}$ can be achieved.  The variations contain contributions from both cluster inhomogeneities and the assumption that the sources were all at midplane, when actually they were sampling only the pathlength appropriate to their 3D position in the cluster and the particular projected view.

\subsection{Comparison to current cluster analyses}
\label{compare}

One of the best data sets for cluster RM analysis comes from Hydra A, the central radio source in a cool core cluster, mapped by \cite{hydra1}, and further analyzed by \cite{vogt04} and \cite{Bayesian1}.  The latter work presents a Bayesian analysis of the northern lobe of the source, which extends $\sim$40 kpc from the center, comparable to the inner core radius of the X-ray distribution.  They derive the magnetic field power spectrum, finding a portion with slope $\sim \frac{5}{3}$,  a characteristic RM coherence length of $\sim$5~kpc and a central field strength of $7\pm2 \mu$G .  Although they describe these errors as reflecting the systematic uncertainties in the distance along the line of sight (reflected in the inclination of the radio source) and in the range in acceptable values of $\eta$ from 0.1 to 0.8, results are only reported holding one or the other of these parameters fixed.  Their equivalent $\Lambda_0$ is held fixed, and since the full spatial resolution is only utilized over $\sim$20~kpc, because of signal:noise concerns, there is no information available on any radial variation in the coherence length.  However, the most important shortcoming of this analysis is that it applies only to the northern lobe.  The southern lobe was explicitly excluded from this analysis because it has different RM properties, with a much stronger RM power spectrum, leading to RM values reaching $\sim$12,000 $\frac{rad}{m^2}$, with associated strong depolarization. { A later analysis of Hydra~A, including both lobes and assuming that the lobes had created cavities in the surrounding medium, was performed by \citet{laing08}.  The latter authors find a) less gas, b) a different preferred value for $\eta$, c) a magnetic field autocorrelation length twice as high, and d) a central magnetic field strength 2.5 $\times$ higher than \citet{Bayesian1}.  The characterization of the central magnetic field strength with high accuracy in \citet{Bayesian1} therefore does not reflect the uncertainties which are present, as shown in this paper, in the modeling of the magnetic field in this cluster.}

\cite{Abell2199} use a different type of analysis to measure the magnetic power spectrum in Abell 2199.  Their approach is also Bayesian, and involves comparing a range of simulations with the observed RM distributions of 3C338, a central radio galaxy with RM measurements extending over $\le 40$kpc.  The maximum fluctuation scale is only approximately characterized at $35\pm28$ kpc, which is expected given the limited RM sampling available. They explicitly show the sensitivity of the central magnetic field strength to the maximum scale, as we also discuss here.  Their final estimates of $\eta$ are from $0.4~-~1.4$, and a central magnetic field strength of $2.7-20.7$.  This large range is consistent with the results of our experiments. 

A study of Abell 194 using RM observations from the Sardinia Radio Telescope and the VLA are presented by \cite{A194}. They use a Bayesian type analysis based on three-dimensional simulations assuming a power law magnetic field spectrum, while utilizing both the RM distribution and fractional polarization information. They find $\eta = 1.1 \pm 0.2$, a maximum scale of $64 \pm 24$ kpc, and a central magnetic field of $1.5 \pm 0.2 \mu$G.  Their polarization information extends over 260~kpc, so we would expect the variations in the field to be very well sampled.  However, the observed RM distribution has a mean value of 15.2 $\frac{rad}{m^2}$ with $\sigma_{RM}$=14.4 $\frac{rad}{m^2}$.   {In this case,  $|\langle RM\rangle|/\langle\sigma_{RM}\rangle$ is not small, as required by the analytical modeling described above, and only a Monte-Carlo type analysis such as performed by \citet{A194} can be used.  In A194, most of the polarized emission comes from 3C40B, associated with a luminous galaxy in the cluster core, so the position along the line of sight is reasonably well constrained. Remaining uncertainties, such as discussed in the \cite{A194} analysis, include possible non-power law magnetic field distributions and/or cavities produced by the radio galaxy lobes \citep[e.g.,][]{laing08}. }

In summary, we find that existing derivations of cluster fields from RM observations reflect the same kinds of fundamental uncertainties that underlie our analysis.  In some published observational analyses, the derived central field strengths are reported with uncertainties consistent with our findings, while in other cases, the extent of the uncertainties is not adequately addressed, and may be seriously underestimated.


\section{Conclusions}

The derivations of central magnetic field strengths in clusters of galaxies are subject to a number of important uncertainties.  From a physical standpoint, cluster magnetic fields are likely to be statistically inhomogeneous, being influenced by disturbances due to the continuing growth and evolution of the cluster, even when no major cluster encounters have happened recently.  Then, even in this ``best case'' situation, a specific ``central field value'' may not be appropriate or adequate to address key science questions associated with the ICM. This needs to be evaluated on a case-by-case basis.

Using as a test base the known magnetic field distribution in a relatively quiescent cluster formed during cosmological simulations, we have found that magnetic field determinations, even in the perfect, but unrealizable case of a fully sampled RM distribution, are limited to a range of $\sim$3.  In the case of  actual observations of real clusters, several practical, irreducible limitations can cause estimates of the central field strengths to span at least an order of magnitude.  These limitations include the unknown scaling of magnetic field strength with ICM electron density, RM sampling limitations that lead to underestimates of the magnitude of RM fluctuations, and possible, but unknown variations of magnetic field structure scales with distance from the cluster center. These limitations in obtaining fits to these model  parameters and the desired magnetic field strengths are amplified by uncertainties such as the unknown positions along the line of sigh of polarized sources used to determine RMs. It will thus always be necessary to introduce both physical and sampling assumptions into any derivations of cluster fields.  It is important that all such assumptions are clearly stated, and that the uncertainties in those assumptions be reflected in the final derived values.


\begin{acknowledgements}

We thank Paul Edmond and the CfA-ITC for providing computation time allowing us to run some of our simulations.
Support for this work has been provided by NSF grants AST-1211595, AST-1714205, NASA grant NNX09AH78G and the Minnesota Supercomputing Institute for Advanced Computational Research.
We thank Jean Eilek and Robert Laing for useful discussions pertaining to this work.

\end{acknowledgements}
\clearpage
\section*{Appendix A}
\label{app:A}
 {In this Appendix, we discuss the issues surrounding the use of the RM coherence length ($\Lambda_0$ and $\Lambda(a)$) needed for the derivation of $\sigma_B$.}  For a magnetic field that is isotropically disordered through turbulence, there will be some effective RM coherence scale along a line of sight of length $\ell$, where $\Lambda  \ll \ell$.  It has been shown that for an isotropic, turbulent field distribution the length $\Lambda$ needed in Eqn. \ref{eq:sigmaRM} can be expressed in terms of the magnetic field power spectrum and the so-called ``integral length'', $L_{int}$, as \citep{ClusterLengths}
\begin{align}
 \Lambda = \frac{3}{4}{L_{int}} = \frac{3}{4} \frac{\int{P_{B}(k) / k \, \mathrm{d}k}}{\int{P_{B}(k)  \mathrm{d}k}}.
\label{eq:intlength}
\end{align}
This relationship depends only on isotropy of the magnetic field and a well-defined magnetic field power spectrum. 
One can alternatively write $\Lambda = 3/2  L_B$, where $L_B$, as defined, for example, in \citet{Bayesian2} represents the magnetic field coherence length. 
In Eqn. \ref{eq:intlength} $P_B(k)$ is the 1-D power spectrum of the 3D, isotropic magnetic field, with the wave number, $k$, defined without the usual $2\pi$ factor.

The numerical factors $3/4$ and $3/2$ connecting $\Lambda$ to $L_{int}$ and $L_B$ reflect different weights that turbulent magnetic field fluctuations contribute to the specific, statistical, length measures. 
Indeed, a number of authors have emphasized in realistic models of RM properties associated with disordered magnetic fields that while $\Lambda \propto L_B$ (or similarly $\Lambda \propto L_{int}$), $\Lambda \ne L_B$ (or, similarly $\Lambda \ne L_{int}$). 

In practice, $\Lambda$  must be estimated from the RM distribution's 2D characteristic coherence scale.  {The observed RM distribution is a projected rather than a local measure. Although the projected RM $\Lambda$ is related to the 3D magnetic field measure, $\Lambda(\vec{r})$, it is not generally equivalent.}  As pointed out above and emphasized by \citet{Bayesian2},  {in a homogeneous isotropic magnetic field setting $\Lambda$ can be directly related to the magnetic field correlation length, $L_B$, by the RM distribution power spectrum \citep[cf. also][]{vogt03}, which can, in principle, be established from observations.} 
However, obtaining both $L_B$ and $\Lambda$ in this way is a complex and difficult procedure in practice,
especially with restricted RM sampling \citep{Bayesian1}. So, a more common strategy has been to assume for modeling purposes a magnetic field power spectrum, usually a power law  with inner and outer scales \citep[e.g.,][]{obsmodel1, espinosaFR2}. On the other hand, MHD simulations reveal that magnetic field distributions evolved through the turbulent dynamo from a weak seed
field are poorly represented by power law power spectra \citep[e.g.,][]{ClusterLengths,porter15}. That is also apparently the case for the magnetic fields that evolve in clusters formed in cosmological simulations \citep[e.g.,][]{xu11,wittor16}, which certainly is the case for the simulated cluster we use in these experiments (\S 2, Figure \ref{fig:PS}). Consequently, we apply in our work here a relatively simpler
approach to estimating $\sigma_{B}$ that does not require assumptions about or computation of power spectra. In particular, as discussed immediately below, we estimate a projected RM coherence length from the 2$^{nd}$ order RM structure function, $\Lambda_{SF}$,
and associate that length with $\Lambda$ in Eqn. \ref{eq:sigmaRM} when deriving magnetic field distribution properties (\S 3.1).

\subsection*{A.1~ Estimating $\Lambda_0$ from RM Data}
The central RM coherence length, $\Lambda_0$, is probably the most challenging observational measure needed to estimate $\sigma_{B,0}$.  {Under the assumption of a power law for the magnetic field (and RM) spectra, one can first optimize the fits to the slope and inner and outer scales of the RM spectrum}  \citep[e.g.,][]{obsmodel1,Bonafede10,Abell2199}. Then the effective $\Lambda$ that applies to Eqn.  \ref{eq:sigmaRM} is a fraction of the outer scale used in the power
law, with the exact value depending on the slope of the power law. 

 {This approach is not valid for non-power-law spectra, and specifically not for the cosmology-derived cluster considered here.  We therefore adopt a different approach,} estimating the RM coherence length, $\Lambda$, from computed $2^{nd}$ order structure functions of the RM distribution over the observed
surface without
assuming any particular form for the magnetic field power spectrum. 

The needed structure function, $S(|\vec\Delta a|)$,  is given by
\begin{align}
S(|\Delta \vec a|) = \langle (RM(\vec a) - RM(\vec a + \Delta \vec a))^{2} \rangle.
\label{eq:RM-sf}
\end{align}
where $\Delta \vec a$ measures an offset, or ``lag'' relative to a specific $\vec a$. The results are then averaged over some specified area where RMs are available.
Meaningful estimates for $S(\vec{a},|\Delta \vec{a}|)$ require the averaging to be done over an area  spanning scales larger than $\Lambda$, which
is also true for estimates of $\sigma_{RM}$, as discussed above.

Mathematically, $S(|\Delta\vec{a}|)$ is just twice the difference between $\langle (RM(\vec a))^2\rangle$ and the RM autocorrelation, $\langle RM(\vec a) \times RM(\vec a + \Delta \vec a)\rangle$, over the defined area.
Then  $S(|\Delta \vec{a}|)$ should approach zero for small lags, while for  $|\Delta \vec{a}|$ larger than the RM correlation length it should approach  $2\langle (RM)^2\rangle$. For our simulated cluster, $\mid\langle RM \rangle\mid\ll \sigma_{RM}$ for suitably large areas (see Figure \ref{fig:etaBICs}), so $S(|\Delta \vec a|)$ should approach 2$\times\sigma_{RM}^2$ for large  $|\Delta \vec a|$. As detailed below, we designate the length $|\Delta a|$ on which $S(|\Delta \vec a|)$ reaches its maximum as $\Lambda_{SF}$, which we then use that as our estimate for $\Lambda$.

We should keep in mind, of course, that while our simulated ICM provides information on the necessary scales, actual cluster RM observations may not provide sufficient sampling, and those approximations may not be meaningful. { However, attempts have been made to directly fit the observed structure functions to high quality RM images by using the Hankel transform \citep[e.g.,][]{laing08}.}

\begin{figure}  
 \includegraphics[height=.8\textheight]{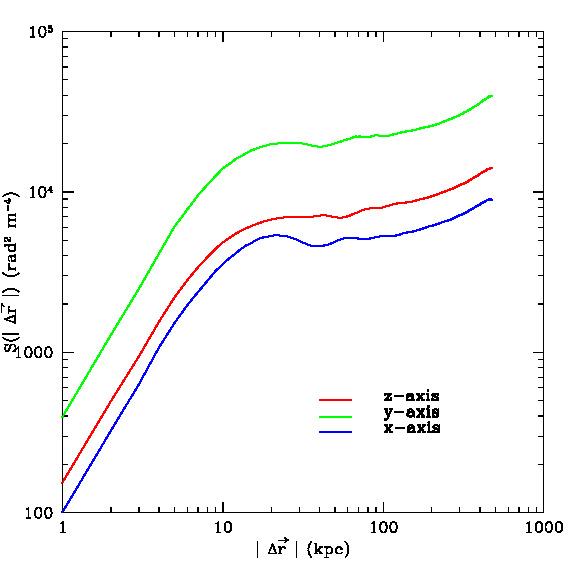}
   \caption{$2^{nd}$ order structure functions of the measured  RM distributions for the {\textit{ g676}} ICM across the full, (1 Mpc)$^2$ analysis box seen down the z-axis (red), y-axis (green) and  x-axis (blue). 
   }
   \label{fig:rmsffullgrid}
\end{figure}

The behavior of $S(|\Delta \vec{a}|)$ computed using background screens spanning {\it {our full $(1~\rm{Mpc})^2$ analysis grid}} in the {\it{g676}} cluster is shown in Figure \ref{fig:rmsffullgrid}. For each projection,  $S(|\Delta \vec a|)$ increases from small scales, then plateaus near values, $S(|\Delta \vec{a}|) \sim 5000 - 10000~{\rm{rad^2/m^4}}$, corresponding to (cluster-wide) $\sigma_{RM} \sim 50-70~{\rm{rad/m^2}}$. Note that this value is an order of magnitude below $\sigma_{RM,0}$, such as values in Figure \ref{fig:etaBICs}, because it represents an average over the entire  $(1~\rm{Mpc})^2$ grid, and is dominated by the very weak fields far from the cluster center.  Nonetheless, the shape of  $S(|\Delta \vec{a}|)$ is a good representation of the magnetic field structure, so can be used to estimate the coherence length.

We define  $|\Delta \vec{a}|_m$ as the smallest lag for which $\frac{\mathrm{d}(\mathrm{ln}(S(\vec{a},|\Delta \vec{a}|)))}{\mathrm{d}(\mathrm{ln}(|\Delta \vec{a}|))} = 0$, and justify this now as our estimated $\Lambda$.
Applied to the full $(1~\rm{Mpc})^2$ analysis box (Figure \ref{fig:rmsffullgrid}) from an average of the three principal axis orientations, we would obtain in this way $|\Delta \vec{a}|_m =  \Lambda \approx 25$ kpc.
A similar analysis of the $S(|\Delta \vec{a}|)$ distributions computed over  $(100~\rm{kpc})^2$ areas centered on the cluster core results in a coherence scale, $|\Delta \vec{a}|_m  = \Lambda \approx 20$ kpc.

Eqn. \ref{eq:BRMSeval} specifically requires an estimate of the central coherence length, $\Lambda_0$. So
it is necessary to establish the relationship between observationally derived $\Lambda$ values  over a selected projected  cluster area and the central $\Lambda_0$. If one assumed a constant $\Lambda$ in evaluating the RM structure function data using fully sampled regions spanning the entire, (1 Mpc)$^3$, analysis box from cluster {\textit{g676}}, they would obtain $\Lambda_0 = \Lambda \approx 25$kpc. The fully sampled core alone would yield  $\Lambda_0 = \Lambda \approx 20$ kpc.
These two  $\Lambda_0$ estimates are similar enough that, taken at face value, they
would lead to estimates for $\sigma_B$ from Eqn.\ref{eq:RM-model1} differing by only roughly 10\%. However, that they do differ points back to the previously discussed issue that {\it the magnetic field integral length is not a constant  value, but increases with distance from cluster center (that is, $L_{int} = L_{int}(r)$, with $dL_{int}/dr > 0$. As pointed out previously, this can have a substantial impact on the derivation of  valid estimates for $\sigma_{B,0}$, depending on how the estimates are made.} 

Specifically, our analysis in \S 2 of the 3D magnetic field properties of {\it{g676}}
established  that $\Lambda = (3/4) L_{int} \approx 17$ kpc  in the central $100~\rm{kpc}^3$ box (so $r\sim 50$ kpc). Since this ($50$ kpc)$^3$ volume is  roughly the size of the core,
we, therefore, obtain directly from the field itself  (not via RM measurements) the estimate $\Lambda_0 \approx 17$ kpc. On the other hand, we similarly found  $\Lambda =(3/4)L_{int} \approx 40$ kpc within the full ($1~\rm{Mpc})^3$ box (so $r \sim 500$ kpc). That is an increase from the core to the outskirts of the cluster of a factor $\sim 2.4$, representing a difference of $\sim 50$\% in the translation between $\sigma_{B,0}$ and $\sigma_{RM,0}$ in Eqn. \ref{eq:BRMSeval}. This enhanced impact comes from the fact that the RM values depend on both the magnetic field distribution and the electron density distribution. The electron density distribution has a steep radial dependence ($\sim r^{-9/4}$ at large radii), which leads to a strong central bias in RM contributions.
Thus, substantially better estimates for $\Lambda_0$ should come from application of the nonuniform model represented in  Eqn. \ref{eq:RM-model2} than from fixed values
for $\Lambda$ represented in Eqn. \ref{eq:RM-model1}.

\subsection*{A.2 Some Power Spectrum Issues}
As emphasized in the discussion after Eqn. \ref{eq:sigmaRM}, the RM dispersion, $\sigma_{RM}$, for a medium with an isotropically disordered magnetic field scales over long paths as the square root of the number of independent magnetic structures along the path, so depends inversely on the square root of $\Lambda$, which scales with the magnetic field  integral correlation and correlation lengths, $L_{int}$ and $L_B$, according to Eqn. \ref{eq:intlength}.
As emphasized in the main body of this paper, $\Lambda$ can only be estimated observationally from
measurements of the RM distribution. We can, however, in our current experiments, compare those observational estimates to the actual $L_{int}$ of the 3D magnetic field. 
Figure \ref{fig:PS} shows the 1D magnetic field power spectrum, $P_B(k)$,  inside our  (1 Mpc)$^3$ analysis box (blue line). For comparison we also show the {\it{g676}} ICM kinetic
energy 1D power spectrum in the same volume (red line), along with a $k^{-5/3}$ line (black), representing the slope of a Kolmogorov spectrum. Here, $k = 1$ corresponds to  a length of 1 Mpc. 

To facilitate  our discussions below
we note some additional relevant properties of these power spectra. The kinetic energy power spectrum in Figure \ref{fig:PS} is very
roughly consistent with the $k^{-5/3}$ form. However, as already pointed out, the broadly peaked magnetic power spectrum is not at all well-represented by a power law. 

From the form of Eqn. \ref{eq:intlength} it is obvious that $L_{int} \sim k_{peak}^{-1}$. Over the full (1 Mpc)$^3$ box, we therefore see from Figure \ref{fig:PS} that $L_{int} \sim 50$ kpc.  Indeed, applying the power spectrum 
in Figure \ref{fig:PS}
we obtain numerically $L_{int}$ = 54 kpc.  On the other hand, as mentioned previously, we generally expect $L_{int}$ to vary with the scales of ICM structures. 
For instance, flux freezing during compression of a disordered magnetic field would lead to $L_{int} \propto \ell \propto n_e^{-1/3}\propto B^{-1/2}$. The exact scaling would depend on dynamical circumstances analogous to the scaling for $B$ itself. Nonetheless, we would expect from dynamical ``similarity'' arguments that $L_{int}$ would usually increase towards the cluster outskirts \citep[e.g.,][]{shi18}. As discussed in the main text, we express this behavior in terms of a density scaling as a convenient proxy. 
We might guess a generalization of our
flux-freezing example to take a form something like $L_{int}\propto n_e^{-\eta/2}$. Since the scaling is weak, the detailed form should not be critical,
and this has the modeling advantage of not adding free parameters to data fits. To test this expectation in the \textit{{g676}} ICM magnetic field we computed magnetic power spectra in nine additional cluster-centered volumes spanning scales ranging from 100 kpc to 900 kpc. Indeed we found that $L_{int}$ increased smoothly from $L_{int} \approx 23$ kpc in the smallest box (essentially including only the cluster core)  to the aforementioned $L_{int} \approx 54$ kpc in the $1~\rm{ Mpc}^3$ box.  The values fit a tight correlation $L_{int} \propto \langle \, n_{e} \, \rangle^{-1/4}$, as well, (not shown) so consistent with our guess that $L_{int} \propto n_e^{-\eta/2}$. We will find below that results of our RM
analysis are substantially improved if we allow $L_{int}$ to increase with distance from the cluster center using this scaling compared to keeping
$L_{int}$ fixed throughout the cluster. \ed{Such a radial dependence would be suggested by the change in the derived slope of the power law spectrum in A2255 as a function of distance from the cluster center \citep{RandomLOS}.}


\clearpage



\begin{deluxetable}{cccc|cccc}
\rotate
\tablecaption{Statistical RM estimates of the ICM magnetic field using background (BG) screens 
}
\tablehead{
   \colhead{}
   & \colhead{$\langle \Lambda_{0} \rangle$}
      & \colhead{$\langle \sigma_{B,0}\rangle$}
 & \colhead{$range(\frac{\sigma_{B,0}}{ \langle\sigma_{B,0} \rangle})$}  \vline
   & \colhead{$\langle \sigma_{B_{0}} \rangle$}
    & \colhead{$range(\frac{\sigma_{B,0}}{ \langle\sigma_{B,0} \rangle})$}
   & \colhead{$\langle \eta \rangle$}
   & \colhead{$range({\eta})$}\\
   \colhead{}
   &\colhead{kpc}
   &\colhead{$\mu$G}
   &\colhead{} \vline
   &\colhead{$\mu$G}
   &
   &\colhead{}
   &\colhead{}
}
\startdata
&&Fixed $\eta = 0.5$&&&&Fitted $\eta$\\
$\Lambda = \Lambda_0$&&&&$\Lambda = \Lambda_0$&&\\
BG Annuli &22& 3.33 & 0.96-1.07 & 2.15 & 0.62-1.70 & 0.38 & 0.31-0.50\\
$\Lambda = \Lambda(a)$&&&&$\Lambda = \Lambda(a)$&&\\
BG Annuli &16 & 1.96 & 0.94-1.06 & 2.30 & 0.61-1.72 & 0.51 & 0.41-0.67\\
\enddata
\tablecomments{Averages and errors are calculated including all three viewing directions for a given experiment. 
The $\eta$ parameter defines the scaling $\sigma_B \propto n_e^{\eta}$. The assumed form for $\Lambda(a)$ is given in Eqn. \ref{eq:varlamb}.
The $\sigma_{B,0}$ values on the left of the table assumed $\eta = 0.5$, while  $\sigma_{B,0}$ values on the right correspond to the associated, fitted $\langle\eta\rangle$, with central electron density, $n_0 = 4\times 10^{-2}$cm$^{-3}$ and ICM core radius, $r_c = 41$ kpc.}
\tablecomments{The fit to the actual 3D magnetic field in the cluster yields $\sigma_{B_{0}} = 1.9 \mu G$ and $\eta = 0.5$.}
\label{tab:BICs}
\end{deluxetable}

\begin{deluxetable}{cccccc}
\tablecaption{Statistics for randomly embedded masks.
}
\tablehead{
   \colhead{}
   & \colhead{$\langle \Lambda_{0} \rangle$}
   & \colhead{$\langle \sigma_{RM,0} \rangle$}
   & \colhead{$trimmed~ (\frac{\sigma_{RM,0}}{\langle\sigma_{RM,0} \rangle})$}
   & \colhead{$\langle \sigma_{B,0} \rangle$}
   & \colhead{$trimmed~ ( \frac{\sigma_{B,0}}{\langle \sigma_{B,0} \rangle})$}\\
   \colhead{}
   &\colhead{kpc}
   &\colhead{rad m$^{-2}$}
   &\colhead{}
   &\colhead{$\mu$G}
   &\colhead{}
}
\startdata
3 Sources & 16 & 526 & 0.71-1.2 & 1.5 & 0.73-1.2\\
8 Sources & 21 & 462 & 0.74-1.2& 1.2 & 0.75-1.2\\ 
\\ 
\enddata
\begin{flushleft}\tablecomments{Averages include all six viewing directions for a given experiment.
$\sigma_{B,0}$ is computed from $\sigma_{RM,0}$  using Eqn. \ref{eq:BRMSeval} and \ref{eq:varlamb}, with $\eta = 0.5$, $\Lambda_0 = \langle\Lambda_0\rangle$, central electron density, $n_0 = 4\times 10^{-2}$cm$^{-3}$, ICM core radius, $r_c = 41$ kpc and assuming the sources are all in the cluster mid-plane. \textbf{Trimmed ranges exclude the 2  most extreme values, so would represent 67\% probabilities if the distributions were Gaussian.  Full ranges for these experiments and those reported in Table 3 were typically an order of magnitude or more.}}
\end{flushleft}
\label{tab:passsource}
\end{deluxetable}

\begin{deluxetable}{cccc}
\tablecaption{
Fits  from Passive AGN Masks.  
\label{tab:Bjet}}
\tablehead{
   \colhead{}
   & \colhead{$\langle \Lambda_{0} \rangle$}
   & \colhead{$\langle \sigma_{B,0} \rangle$}
   & \colhead{$trimmed~ (\frac{\sigma_{B,0}}{\langle \sigma_{B,0} \rangle})$}\\ 
   \colhead{}
   &\colhead{kpc}
   &\colhead{$\mu$Gauss}
   &\colhead{}
}
\startdata
 t =79 Myr& 16 & 1.95 & 0.72-1.7 \\
t = 92 Myr & 16 & 1.80 & 0.63-1.5\\
\\
\enddata
\tablecomments {Each $\langle\sigma_{B,0}\rangle$  is derived from RM data along eight viewing directions using the same fitting parameters as for values in Table \ref{tab:passsource}.  The physical, 3D ICM magnetic field properties evolved to the observation time 79 Myr (92 Myr) were $\sigma_{B,0} = 1.79 ~\mu$Gauss, $\eta$ = 0.41 ($\sigma_{B,0} = 1.60~ \mu$Gauss, $\eta$ = 0.39). See Table 2 notes regarding trimmed values.}
\label{tab:agnmasks}
\end{deluxetable}

\clearpage

\begin{figure}  
\begin{center}
  \includegraphics[height=.8\textheight]{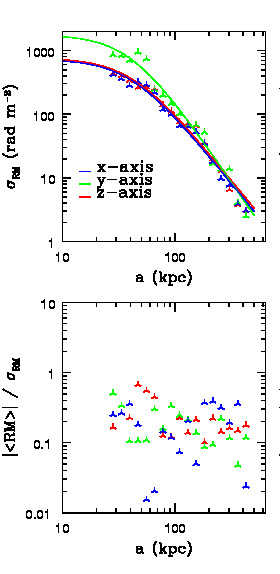}
\end{center}
    \caption{Distributions and fits to the {\it background screen} $\sigma_{RM}$ for: {\bf (Upper panel:)} logarithmically spaced, circular annuli of projected radius, $a$, (10 kpc thickness) for views along each principle axis of the (1 Mpc)$^3$ {\textit{ g676}} analysis box,
See Table \ref{tab:BICs} for fitting summaries. {\bf  Bottom panel:}   normalized means, $|\langle RM \rangle| / \sigma_{RM}$ corresponding to $\sigma_{RM}$ results in the  {upper} panel 
Statistical errors for the annuli data points are symbol sized or smaller.  
} 
    \label{fig:etaBICs}
\end{figure}

\begin{figure}  
   \includegraphics[height=.8\textheight]{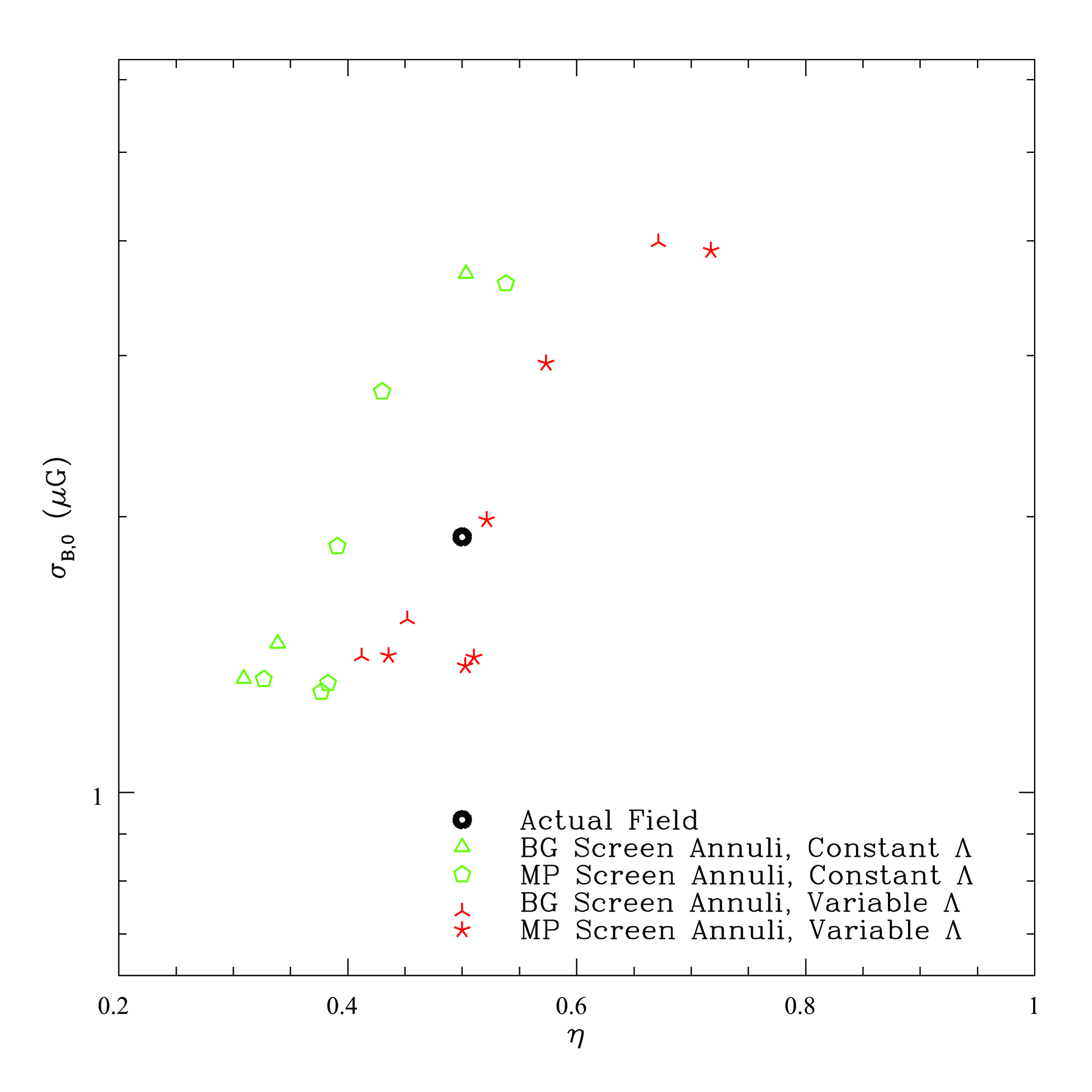}
   \caption{Empirically estimated $\sigma_{B,0}$ and $\eta$ using  $\sigma_{RM}$ calculated from  mid-cluster (5-point symbols) and background (3-point symbols) annular screens.
Constant coherence scale, $\Lambda$, solutions are green, while density dependent $\Lambda$ solutions are red.  The black circular point marks the actual, 3D {\textit {g676}} global magnetic field properties.}
   \label{fig:etaBIC2s}
\end{figure}

\begin{figure}  
   \includegraphics[height=.62\textheight]{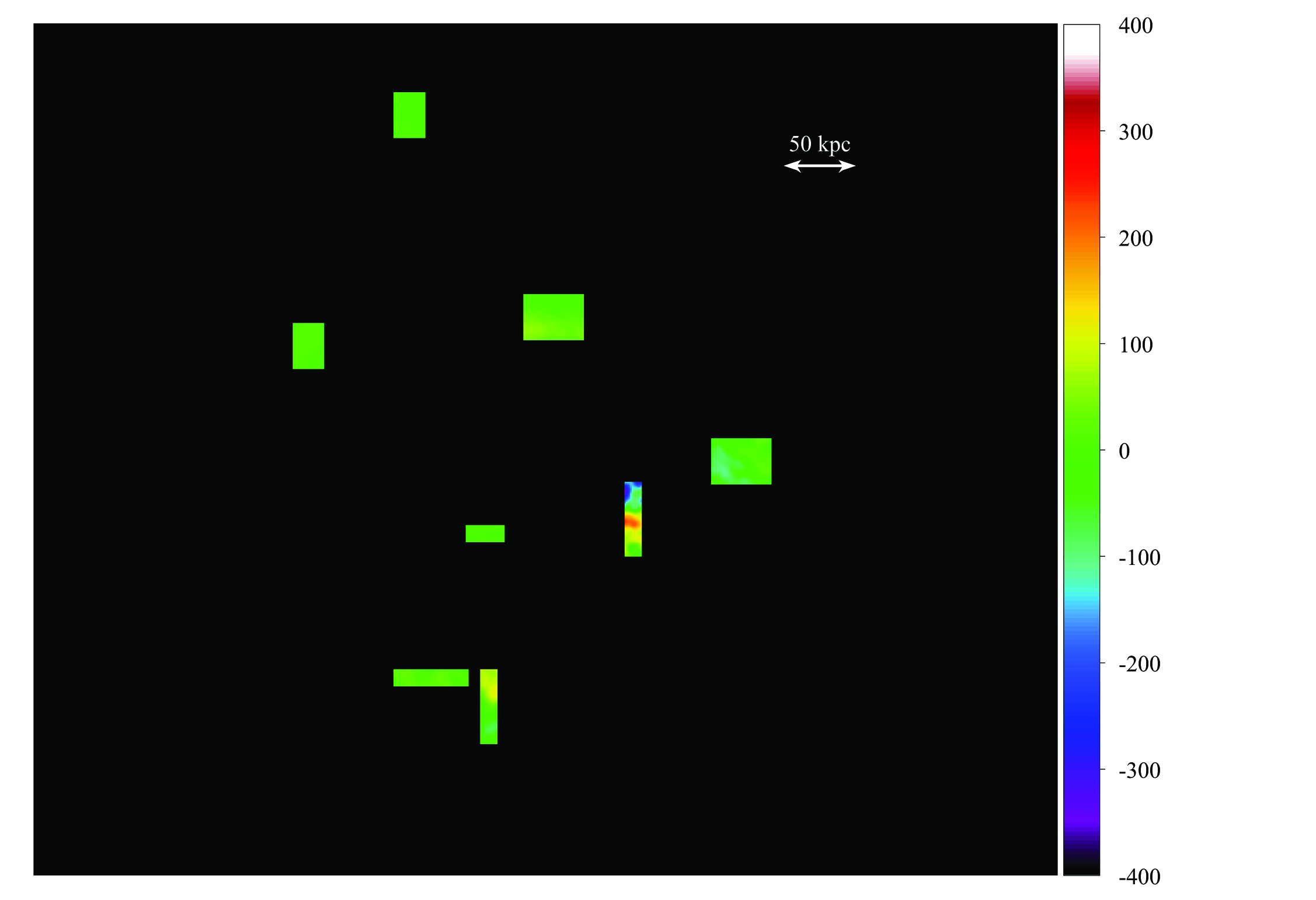}
   \caption{RM map for 8 randomly distributed, embedded but passive ``source'' screens viewed down the z-axis of the {\textit{g676}} analysis box. The projected cluster center is in the middle of the view. Individual source projected distances from the near box face, $a_i$, range from $\sim$ 70 kpc to $\sim$ 250 kpc. The colorbar unit is $\mathrm{rad \, m^{-2}}$.}
   \label{fig:randsource}
\end{figure}

\begin{figure}  
 \includegraphics[height=0.8\textheight]{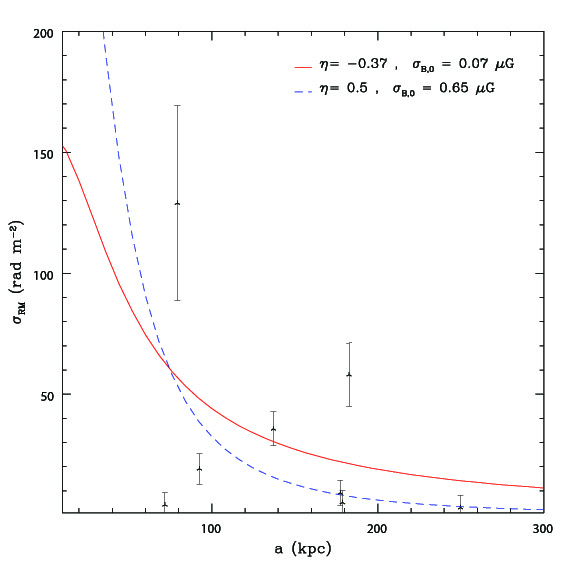}
   \caption{RM statistics from the  passive, ``embedded source'' screens shown in Figure \ref{fig:randsource}. Two fits for $\sigma_{RM}(a)$ as described in \S 4.2 are shown. The solid (red)  curve includes $\eta$ as a free parameter, while the dashed (blue) curve fixes $\eta$ = 0.5. The intercept of the
latter fit (off scale) is $\sigma_{RM,0} = 400 {\rm rad~m}^{-2}$. The derived values for $\sigma_{B,0}$ were then obtained using $\Lambda_0 = 32$kpc, taken from the
structure functions of the associated  RM distribution,  assuming, $r_c = 41$kpc, $n_0 = 0.0378 {\rm cm}^{-3}$. Error bars on the individual measurements are underestimated, based on an optimistic assumption of the number of independent points in each mask; however, the actual scatter among them is much larger than any statistical uncertainties, and arise from the actual variations along the different lines of sight to the mask.} 
   \label{fig:sigfit}
\end{figure}

\begin{figure}  
     \includegraphics[height = 0.79\textheight]{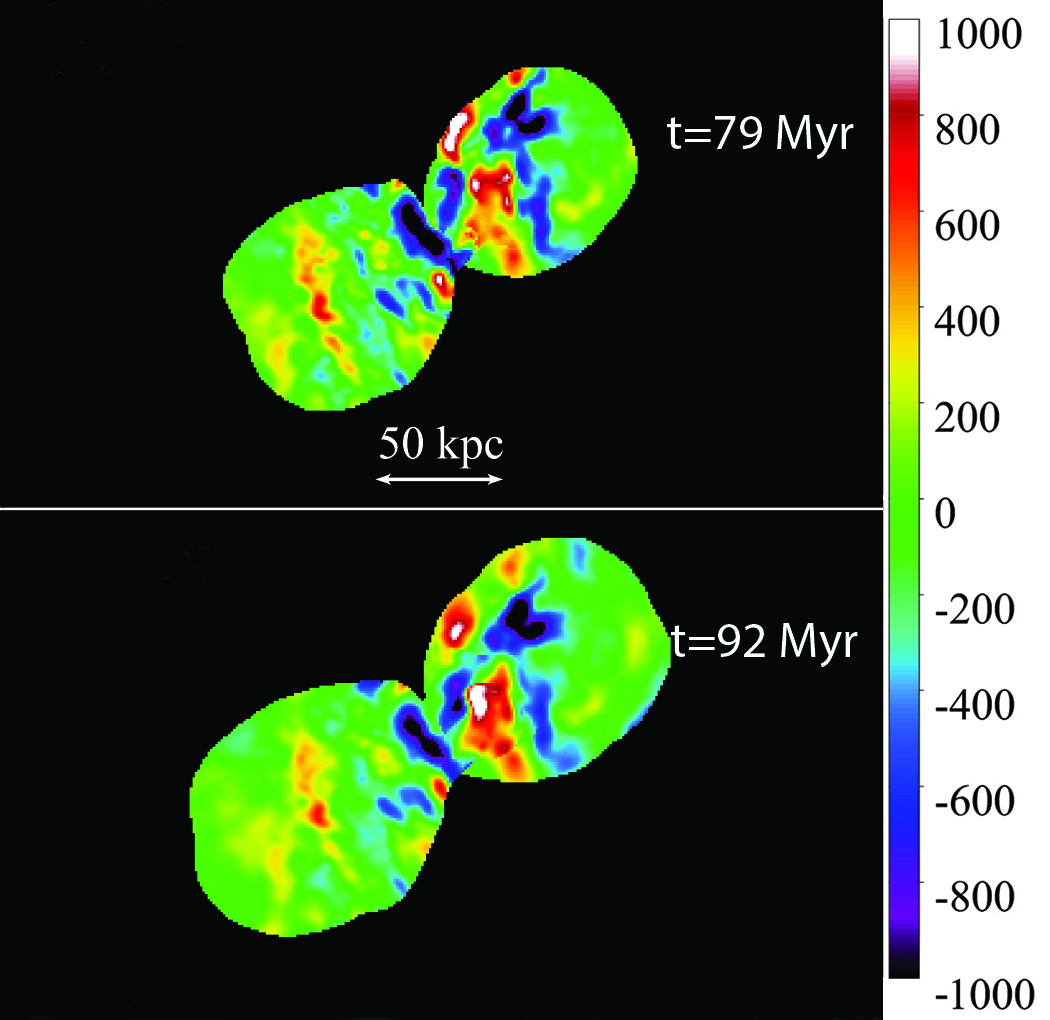}
   \caption{RM maps of the embedded, passive bipolar AGN masks viewed down the grid z-axis at t = 79 Myr (top) and t = 92 Myr (bottom) Colorbar units are rad/m$^2$.
   The AGN jets are active at the lower time, but inactive at the upper time.
   Details of the mask construction are in the text.}
   \label{fig:rmjetmap}
\end{figure}

\begin{figure} 
  \includegraphics[height=.8\textheight]{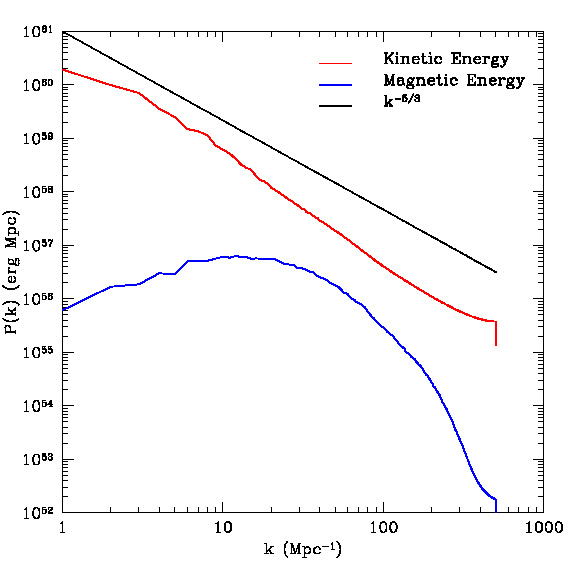}
   \caption{Power Spectrum of the kinetic energy (red) and magnetic energy (blue) of the {\textit{g676}} ICM within the (1 Mpc)$^3$ analysis box.  Solid, black line
represents a Kolmogorov spectrum, $P(k) \propto\mathrm{k^{-5/3}}$, spectrum.  The broad magnetic power spectrum turnover,
$\mathrm{k \sim 8- 20 ~Mpc^{-1}}$,  represents scales $\mathrm{k^{-1} \sim 50 - 120 ~kpc}$. 
}
\label{fig:PS}
\end{figure}


\begin{thebibliography}{9}

\bibitem[Ascasibar \& Markevitch(2006)]{ascasibar06}
Ascasibar, Y. \& Markevitch, M., 2006, \apj, 650, 102

\bibitem[Bonafede \etal (2010)]{Bonafede10}
Bonafede, A., Feretti, L., Murgia, M., Govoni, F., Giovannini, G., Dallacasa, D., Dolag, K. \& Taylor, G. B., 2010, A\&A, 513, A30

\bibitem[Bonadede \etal (2013)]{Bonafede13}
Bonafede, A., Vazza, F., B\"ruggen, M., Murgia, M., Govoni, F., Feretti, L., Giovannini, G., Ogrean, G., 2013, \mnras, accepted


\bibitem[Brunetti \& Jones(2014)]{brun14}
Brunetti, G. \& Jones, T. W. 2014, International Journal of Modern Physics D, 23, 1430007

\bibitem[Carilli \& Taylor(2002)]{MagFieldReview}
Carilli, C. L., \& Taylor, G. B., 2002, ARA\&A, 40, 319

\bibitem[Cavaliere \& Fusco-Femiano(1976)]{BetaLaw}
Cavaliere A., Fusco-Femiano R., 1976, A\&A, 49, 137

\bibitem[Cho \& Ryu(2009)]{ClusterLengths}
Cho, J., \& Ryu, D. 2009, \apj, 705, L90

\bibitem[Cho \etal (2009)]{cho09}
Cho, J., Vishniac, E. T., Beresnyak, A., Lazarian, A. \& Ryu, D., 2009, \apj, 693, 1449

\bibitem[Clarke \etal (1999)]{clarke99}
Clarke, T. E., Kronberg, P. P., B\"orhinger, H., 1999, in ``Diffuse thermal and relativistic plasma in galaxy clusters,''
Eds. H. B\"ohringer, L. Feretti,  \& P. Schuecker (Garching), p82

\bibitem[Dolag \etal (2001)] {Dolag01} 
Dolag, K., Schindler, S. Govoni, F.,  Feretti, L. 2001, A\&A, 378, 777D

\bibitem[Dolag \etal (2005)]{sphmag}
Dolag, K., Grasso, D., Springel, V., Tkachev, I., 2005, JCAP, 1, 9

\bibitem[Dolag \etal (2008)]{nontherm}
Dolag, K., Bykov AM, Diaferio A. 2008. Space Science Reviews 134:311???335

\bibitem[Dolag \& Stasyszn(2009)]{ClusterPaper}
Dolag, K. \& Stasyszyn, F. 2009, \mnras, \textbf{398}, 1678

\bibitem[Dolag \etal (2009)]{dolag09a}
Dolag, K., Borgani, S., Murante, G. \& Springel, V., 2009,\mnras, 399, 497

\bibitem[Donnert \etal (2009)]{donn99}
Donnert, J., Dolag, K. \& Lesch, H., 2009,\mnras, 392, 1008

\bibitem[Eilek \& Owen(2002)]{AbellRMs}
Eilek, J. A., Owen F. N., 2002, \apj, 567, 202

\bibitem[En{\ss}lin \& Vogt (2003)]{Bayesian2}
En{\ss}lin T.A., Vogt C., 2003, A\&A, 401, 835

\bibitem[En{\ss}lin \etal (2003)]{RBRebuttle}
En{\ss}lin T.A., Vogt C., Clarke T.E., Taylor G.B., 2003, \apj, 597, 870

\bibitem[Felten(1996)]{Felten96}
Felten, J. E., 1996, in `` Clusters, Lensing, and the Future of the Universe,'' ed. V. Trimble, \& A.
Reisenegger, ASP Conf. Ser., 88, 271

\bibitem[Feretti \etal (1995)]{feretti95}
Feretti, L., Dallacasa, D., Giovannini, G., Tagliani, A. 1995, A\&A, 302, 680

\bibitem[Feretti \etal (1999)]{Abell119}
Feretti L., Dallacasa D., Govoni F., Giovannini G., Taylor G. B., Klein U., 1999, A\&A, 344, 472

\bibitem[Feretti \etal (2012)]{feretti12}
Feretti, L., Giovannini G, Govoni, F. \& Murgia, M. 2012, A\&AR, 20 54

\bibitem[Govoni \etal (2006)]{RandomLOS}
Govoni, F., Murgia, M., Feretti, L., et al. 2006, A\&A, 460, 425

\bibitem[Govoni \etal (2017)]{A194}
Govoni, F. \etal 2017, A\&A, 603, 122

\bibitem[Guidetti \etal (2008)]{guidetti08}
Guidetti, D., Murgia, M., Govoni, F., Parma, P., Gregorini, L., de Ruiter, H.R., Cameron, R. A., Fanti, R.,
2008, A\&A, 483, 699

\bibitem[Guidetti \etal (2010)]{guidetti10}
Guidetti, D., Laing, R. A., Murgia, M., Govoni, F., Gregorini, L, Parma, P., 2010, A\&A, 514, 50

\bibitem[Guidetti \etal (2011)]{bandedRMs}
Guidetti, D., Laing R.A., Bridle A.H., Parma P., Gregorini L., 2011, \mnras, \textbf{413}, 2525

\bibitem[Guidetti \etal (2012)]{AsymRMs}
Guidetti, D., Laing R.A., Croston J.H., Bridle A.H., Parma P., 2012, \mnras, \textbf{423}, 1335

\bibitem[Heinz \etal (2006)]{heinz06}
Heinz, S., Br\"uggen, M., Young, A. \& Levesque, E., 2006, \mnras, \textbf{373}, L65

\bibitem[Huarte-Espinosa \etal (2011)]{espinosaFR2}
Huarte-Espinosa, M., Krause, M., \& Alexander, P. 2011, \mnras, \textbf{418}, 162


\bibitem[Jones \etal (2011)]{MHDTurbulence}
Jones, T. W., Porter, D. H., Ryu, D., \& Cho, J. 2011, Mem. Soc. Astron. It., 82, 588

\bibitem[Kunz \etal (2011)]{RadMag}
Kunz M., Schekochihin A., Cowley S., Binney J., Sanders J., 2011, \mnras, \textbf{410}, 2446

\bibitem[Laing \etal (2008)]{laing08}
Laing, R., R. A., Bridle, A. H., Parma, P., Murgia, M., 2008, MNRAS, 391, 521

\bibitem[Laing \etal (2011)]{obsmodel2}
Laing, R. A., Bridle, A. H., Parma, P., \& Murgia, M., 2008, \mnras, \textbf{391}, 521

\bibitem[Lawler \& Dennison(1982)]{lawler82}
Lawler, J. M. \& Dennison, B., 1982, \apj, 252, 81

\bibitem[Li \etal (2006)]{towerjets}
Li, H., Lapenta, G., Finn, J. M., Li, S., \& Colgate, S. A. 2006, \apj, 643, 92


\bibitem[Mendygral \etal (2012)]{mendygraljets}
Mendygral, P. J., Jones, T. W., \& Dolag, K., 2012, \apj, 750, 166

\bibitem[Miniati \& Beresnyak (2015)]{miniati15}
Miniati, F. \& Beresnyak, A., 2015, Nature, 523, 59

\bibitem[Murgia \etal (2004)]{obsmodel1}
Murgia M., Govoni F., Feretti L., Giovannini, G., Dallacasa, D., Fanti, R., Taylor, G, B, \& Dolag. K., 2004, A\&A 424, 429

\bibitem[Newman  \etal (2002)]{newman}
Newman, W. I., Newman, A. L. \& Rephaeli, Y., 2002, \apj, 575, 755

\bibitem[O'Neill \& Jones(2010)]{oneill10}
O'Neill, S. M. \& Jones, T. W., 2010, \apj, 710, 180

\bibitem[Parrish \etal (2012)]{aniso}
Parrish I. J., McCourt M., Quataert E., Sharma P. 2012, \mnras, \textbf{422}, 704

\bibitem[Porter \etal (2015)]{porter15}
Porter, D. H., Jones, T. W. \& Ryu, D., \apj, textbf{810}, 93

\bibitem[Rudnick \& Blundell(2003)]{RSContribution}
Rudnick, L. \& Blundell, K. M., 2003, \apj, 588, 143

\bibitem[Ryu \etal (2008)]{ryu08}
Ryu, D., Kang, H., Cho, J. \& Das. S., 2008, Science,  320, 909

\bibitem[Sanders \etal (2011)]{cturb2}
Sanders, J.S., Fabian, A.C., \& Smith, R.K. 2011, \mnras, \textbf{410}, 1797

\bibitem[Schekochihin \etal (2004)]{TurbDynaSchek}
Schekochihin, A. A., Cowley, S. C., Taylor, S. F., Maron, J. L., \& McWilliams, J. C. 2004, \apj, 612, 276

\bibitem[Shi \etal(2018)]{shi18}Shi, X., Nagai, D. \&  Lau, E. T., 2018, \mnras, 481, 1075

\bibitem[Schuecker \etal (2004)]{cturb}
Schuecker P., Finoguenov A., Miniati F., B??ohringer H., \& Briel U. G., 2004, A\&A, 426, 387


\bibitem[Simionescu \etal (2009)]{hydra}
Simionescu, A., Roediger, E., Nulsen, P., Br\"uggen, Forman, W., B\"ohringer, Werner, N. \& Finoguenov, A., 2009, A\&A, 495, 721

\bibitem[Stasyszyn, Dolag, \& Beck (2013)] {dolag13}
Stasyszyn, F. A., Dolag, K.\&  Beck, A. M., 2013, \mnras, 428 13S

\bibitem[Taylor \etal (2006)]{perseus1}
Taylor, G. B., Gugliucci, N. E., Fabian, A. C., Sanders, J. S., Gentile, G., \& Allen, S. W. 2006, \mnras, \textbf{368}, 1500

\bibitem[Taylor \& Perley(1993)]{hydra1}
Taylor, G. B., \& Perley, R. A. 1993, \apj, 416, 554

\bibitem[Tribble(1991)]{tribble91}
Tribble, P. C., 1991, \mnras, 250, 726 

\bibitem[Vacca \etal(2012)]{Abell2199}
Vacca, V., Murgia, M., Govoni, F., Feretti, L., Giovannini, G., Perley, R. A., \& Taylor, G, B, 2012, A\&A, 540, A38

\bibitem[Vazza \etal (2017a)]{vazza17a}
Vazza, F., Jones, T. W., Br\"uggen, M., Gheller, C., Porter, D. H. \& Ryu, D., 2017, \mnras, 464, 210

\bibitem[Vazza \etal (2017b)]{vazza17b}
Vazza, F., Brunetti, G., Br\"uggen \& Bonafede, A. 2017, arXiv:1711.02673

\bibitem[Vikhlinin \etal (2006)]{ClusterProperties}
Vikhlinin, A., Kravtsov, A., Forman, W., Jones, C., Markevitch, M., Murray, S. S., \& Van Speybroeck, L. 2006, \apj, 640, 691

\bibitem[Vogt \& En{\ss}lin (2003)]{vogt03}
Vogt, C., \& En{\ss}lin, T. A. 2003, A\&A, 412, 373

\bibitem[Vogt, Dolag \& En{\ss}lin (2004)]{vogt04}
Vogt, C., Dolag, K. \& En{\ss}lin, T. 2004 . arXiv:astro-ph/0401216

\bibitem[Vogt \& En{\ss}lin(2005)]{Bayesian1}
Vogt, C., \& En{\ss}lin, T. A. 2005, A\&A, 434, 67

\bibitem[Wittor \etal (2016)]{wittor16}
Wittor, D., Vazza, F. \& Br\"uggen, M., 2016. \mnras, \textbf{464}, 4448

\bibitem[Xu \etal (2011)]{xu11}
Xu, H., Li, .H., Collins, D. C., Li, S. \& Norman, M. L., 2011, \apj, 739, 77

\bibitem[Xu \etal (2012)]{XuRMs}
Xu, H., Govoni, F., Murgia, M., Li, H., Collins, D. C., Norman, M. L., Cen, R., Feretti, L., \& Giovannini, G., 2012, \apj, 759, 40

\bibitem[Zhuravleva \etal (2011)]{zhurav_11}
Zhuravleva, I., Chrurazov, E., Sazonov, S., Sunyaev, R. \& Dolag, K., 2011, Ast. Lett., 37, 141

\bibitem[Zuhone \etal (2010)]{zuhone11}
Zuhone, J. A., Markevitch, M. \& Lee, D., 2010, \apj, 743, 16

\end{thebibliography}
\end{document}